\pdfoutput=1
\RequirePackage{ifpdf}
\ifpdf 
\documentclass[pdftex]{sigma}
\else
\documentclass{sigma}
\fi

\numberwithin{equation}{section}

\usepackage[all]{xy}
\usepackage{enumitem}

\newtheorem{tma}{Theorem}[section]
\newtheorem{lma}[tma]{Lemma}
\newtheorem{cor}[tma]{Corollary}
{\theoremstyle{definition}
\newtheorem{dfn}[tma]{Definition}}

\begin{document}

\newcommand{\arXivNumber}{1407.4741}

\allowdisplaybreaks

\renewcommand{\PaperNumber}{048}

\FirstPageHeading

\ShortArticleName{Classical Abelian Theory with Corners}

\ArticleName{General Boundary Formulation for $\boldsymbol{n}$-Dimensional\\
Classical Abelian Theory with Corners}

\Author{Homero G.~D\'IAZ-MAR\'IN~$^{\dag\ddag}$}

\AuthorNameForHeading{H.G.~D\'{\i}az-Mar\'{\i}n}

\Address{$^\dag$~Escuela Nacional de Ingenier\'\i a~y Ciencias, Instituto Tecnol\'ogico y de Estudios Superiores\\
\hphantom{$^\dag$}~de Monterrey, C.P.~58350 Morelia, M\'exico}
\EmailD{\href{mailto:homero.diaz@itesm.mx}{homero.diaz@itesm.mx}}

\Address{$^\ddag$~Centro de Ciencias Matem\'aticas, Universidad Nacional Aut\'onoma de M\'exico,\\
\hphantom{$^\ddag$}~C.P.~58190 Morelia, M\'exico}
\EmailD{\href{mailto:homero@matmor.unam.mx}{homero@matmor.unam.mx}}

\ArticleDates{Received October 30, 2014, in f\/inal form June 04, 2015; Published online June 24, 2015}

\Abstract{We propose a~general reduction procedure for classical f\/ield theories provided with abelian gauge symmetries
in a~Lagrangian setting.
These ideas come from an axiomatic presentation of the general boundary formulation (GBF) of f\/ield theories, mostly
inspired by topological quantum f\/ield theories (TQFT).
We construct abelian Yang--Mills theories using this framework.
We treat the case for space-time manifolds with smooth boundary components as well as the case of manifolds with corners.
This treatment is the GBF analogue of extended TQFTs.
The aim for developing this classical formalism is to accomplish, in a~future work, geometric quantization at least for
the abelian case.}


\Classification{53D30; 58E15; 58E30; 81T13}

\section{Introduction}

In the variational formulation of classical mechanics, time evolution from an ``initial'' to a~``f\/inal'' state in
a~symplectic phase space $(A,\omega)$ is given by a~relation def\/ined by a~Lagrangian space~$L$ contained in the
symplectic product $(A\oplus \overline{A},\omega\oplus-\omega)$.
Similarly classical f\/ield theories can be formalized rigorously in a~symplectic framework.
The evolution relation associates ``incoming'' to ``outgoing'' \emph{Cauchy boundary data} for the case where space-time~$M$ has
incoming and outgoing bounda\-ry components, $\partial M=\partial M_{\rm in}\cup\partial M_{\rm out}$.
Fields are valued along the boundary together with their derivatives.
This relation def\/ines an isotropic \emph{space of boundary conditions that extend to solutions} in the interior of~$M$,
$L_{\tilde{M}}\subset A_{\partial M}=A_{\partial M_{\rm in}}\times A_{\partial M_{\rm out}}$, where the symplectic structure,
$\omega_{\partial M}=\omega_{\rm in}\oplus\omega_{\rm out}$, is formed by certain symplectic structures $\omega_{\rm in}$ and
$\omega_{\rm out}$ def\/ined in $A_{\partial M_{\rm in}}$ and $A_{\partial M_{\rm out}}$, respectively.
For recent progress from a~categorical point of view on this classical formalism in the case of linear symplectic
spaces, see for instance~\cite{We}.
In some cases, degeneracies of the Lagrangian density yield degeneracies of a~presymplectic structure $\omega_{\partial
M}$, for the Cauchy data $A_{\partial M}$.

A wise observation appearing for the f\/irst time in~\cite{KT}, is that it is possible to formulate a~symplectic framework
for f\/ield theories in general space-time \emph{regions}~$M$.
Here, general boun\-da\-ries $\partial M$ are composed of general \emph{hypersurfaces}, which do not necessarily correspond
to ``in'' and ``out'' space-like boundary components.
The spaces $A_{\partial M}$ of $1$-jets arising from Cauchy data, namely, Dirichlet and Neumann boundary data, have
a~presymplectic structure $\omega_{\rm in}$, see~\cite{KT}.
A~derivation of a~symplectic formalism, was independently rediscovered in the \emph{general bounda\-ry formulation} (GBF)
for classical theories in~\cite{O3,O}, this time arising from their quantum counterparts.
Here, the def\/inition of a~(pre)symplectic structure is given for the space $\tilde{A}_{\partial M}$ of germs of
solutions of a~cylinder of the boundary $\partial M_\varepsilon:=\partial M \times [0,\varepsilon]$.
Axiomatic frameworks incorporating this symplectic formalism appeared in~\cite{O3,O}, for linear f\/ield theories, whereas
for the case of af\/f\/ine f\/ield theories they appeared in~\cite{O1}.
Another symplectic setting for f\/ield theories appeared independently in~\cite{CMR}, where it is related to the BFV and
BV formalism.
Here appears explicitly the distinction for the (pre)symplectic structure for $1$-jets and for germs.

The space of germs $\tilde{A}_{\partial M}$ contains much more information than the $1$-jets for f\/ields in~$\partial M$.
As a~consequence, if we consider germs instead of $1$-jets, then instead of a~symplectic structure~$\omega_{\partial
M}$, we may have a~presymplectic structure~$\tilde{\omega}_{\partial M}$.
Hence we need to consider the space of germs of boundary conditions~$\tilde{A}_{\partial M}$ as a~coisotropic space.
This space of germs~$\tilde{A}_{\partial M}$ needs to be reduced in order to obtain a~symplectic space.

We suppose that both degeneracies, those due to germ higher order derivatives as well as those those due to Lagrangian
density, are both contained in the kernel of the presymplectic structure $\tilde{\omega}_{\partial M}$ in
$\tilde{A}_{\partial M}$.
So the reduced space $A_{\partial M}$ is a~symplectic space.

Dynamics in the interior of the space-time region~$M$ is described as a~Lagrangian immersion, $A_{\tilde{M}}\subset
A_{\partial M}$ of the boundary data of solutions of the Euler--Lagrange equations.
For inf\/inite-dimensional symplectic vector spaces, isotropic spaces are required to be coisotropic in order to be
Lagrangian.
Isotropy is always satisf\/ied~\cite{KT}, but coisotropy of the immersion $A_{\tilde{M}}\subset A_{\partial M}$ does not
hold in general, see counterexamples in~\cite{CMR}.

From the quantum side the axiomatic setting for the GBF is inspired by topological quantum f\/ield theories (TQFT),
see~\cite{At} and the approach of G.~Segal~\cite{Seg:cftdef}.
We consider objects in the category of $(n-1)$-manifolds, i.e., closed boundary components or hypersurfaces~$\Sigma$,
provided with additional normal structure required by germs of solutions.
For instance for f\/ield theories without metric dependence we consider gluings by dif\/feomorphisms of tubular
neighborhoods of~$\Sigma$~\cite{Mi}.
Meanwhile, for f\/ield theories depending on the metric we consider gluing by isometries of~$\Sigma$,~$\overline{\Sigma'}$
leaving invariant the metric tensor germ along~$\Sigma$.
The gluing of two regions~$M_1$,~$M_2$ can be performed along hypersurfaces $\Sigma\subset M_1$, $\Sigma'\subset M_2$, both
isometric oriented manifolds,~$\Sigma\cong\overline{\Sigma'}$.
Here~$\overline{\Sigma'}$ means \emph{reversed orientation}.
The precise axiomatic system for quantum f\/ield theories along with their classical counterpart appears in~\cite{O3} and
for af\/f\/ine theories in~\cite{O1}.

{\bf Corners.} This TQFT-inspired approach requires a~classif\/ication of the basic regions or buil\-ding blocks used to
reconstruct the whole space-time region $M_1\cup_\Sigma M_2$, by gluing the pieces~$M_1$,~$M_2$, along the boundary
hypersurface $\Sigma\cong\overline{\Sigma'}$.
This classif\/ication from the topological point of view can be achieved at least for the case of two-dimensional
surfaces.
In higher dimensions, it would be appealing to avoid such classif\/ication issues, by considering simpler building blocks,
such as~$n$-balls.
Unfortunately, the consequence is that we would have to allow gluings of regions along hypersurfaces~$\Sigma$ with
nonempty boundaries $\partial \Sigma$.
For instance, we can consider the gluing of two~$n$-balls~$M_1$,~$M_2$ along $(n-1)$-balls contained in their bounda\-ries~$\Sigma$,~$\overline{\Sigma'}$. This means that we would have to allow non dif\/ferentiability and lack of normal derivatives
of f\/ields along the $(n-2)$-dimensional \emph{corners} contained in the bounda\-ries~$\partial \Sigma$, of boundary faces,
$\Sigma\subset \partial M_1$, $\Sigma'\subset \partial M_2$.
A~well suited language for describing such phenomena, consists in treating regions~$M_i$ as \emph{manifolds with corners}.
For TQFT the attempt to deal with the case of corners gives rise to extended topological quantum f\/ield theories.
A~possible approach for two-dimensional theories is given for instance in~\cite{Ge, LP}.
There is also a~specif\/ic formulation for $2$-dimensional with corners in~\cite{O2}.
Our aim is to extend this last approach to higher dimensions.

\looseness=-1
{\bf Gauge f\/ield theories.} When we consider principal connections on a~principal bundle $P\rightarrow M$, with
structure compact Lie group~$G$, they are represented by sections of the quotient af\/f\/ine $1$-jet bundle
$J^1P/G\rightarrow M$.
In this case the space of sections $K_M$ is an af\/f\/ine space.
Furthermore for quadratic Lagrangian densities we will have that the space of solutions,~$A_M$, is an \emph{affine
space}.
This enables us to consider a~GBF formalism for af\/f\/ine spaces such as is described in~\cite{O1}.

We give a~step further in relation to~\cite{O1} since we consider gauge symmetries,~$G_M$, acting on~$A_M$.
Variational gauge symmetries are vertical automorphisms of the bundle~$P$, that in turn yield vertical automorphisms of
the bundle $J^1P/G$.
Inf\/initesimal gauge symmetries should preserve the action, $S_M\colon K_M\rightarrow \mathbb{R}$.
They can be identif\/ied with vertical~$G$-invariant vector f\/ields $\vec{X}$ on~$P$, as well as with sections of
$VP/G\rightarrow M$, where $VP$ is the vertical tangent bundle of~$P\rightarrow M$.
Action preservation follows from invariance of the Lagrangian density under vertical vector f\/ields act on~$J^1P/G$.

When we consider germs of solutions on the boundary, we have a~group of variational gauge symmetries
$\tilde{G}_{\partial M}$. By taking the quotient by the degeneracies we obtain a~gauge group action~$G_{\partial M}$ of
symplectomorphisms on $(A_{\partial M},\omega_{\partial M})$.
To make sense of the quotient space $A_M/G_M$ may be problematic in non-abelian gauge f\/ield theories, also taking the related reduced boundary conditions
$A_{\partial M}/G_{\partial M}$.
The issue of gluing solutions also needs to be clarif\/ied.

{\bf Main results.} Our aim is to give an axiomatic GBF formulation for gauge f\/ield theories in the case of space-time
regions with corners.
For the classical theory we will consider the following simplif\/ications: Abelian structure groups and af\/f\/ine structure
for the space of solutions to Euler--Lagrange equations.
We use this axiomatic setting to construct abelian theories.
The most general setting of nonabelian structure groups for actions remains a~conjecture even in the classical case,
see~\cite{CMR1}.
Along this program we study the case of smooth space-time regions without corners as well as the case of regions with
corners.

As we were writing this article we realized that Lagrangian embedding for the abelian case of actions and other
important cases such as BF and Chern--Simmons were actually proved in~\cite{CMR1}.
Here the authors use Lorenz gauge f\/ixing and use Dirichlet boundary condition for $1$-forms.
Thus by Friedrichs--Morrey--Hodge theory they describe the space of boundary conditions that extend to solutions modulo
gauge, $A_{\tilde{M}}/G_{\partial M}\subset A_{\partial M}/G_{\partial M}$, as harmonic forms on $\partial M$ extendable
to cocolsed forms on~$M$.
$A_{\tilde{M}}/G_{\partial M}$ is isomorphic to the direct sum of two spaces: On one hand a~f\/inite-dimensional subspace
of $H^1(M,\partial M)$.
On the other hand an inf\/inite-dimensional space of closed forms in $\partial M$, see Proposition~4 in the appendix
of~\cite{CMR1}.
Independently, we use axial gauge f\/ixing and Neumann boundary conditions for $1$-forms to describe the space
$A_{\tilde{M}}/G_{\partial M}$ as a~direct sum of two spaces: On one and a~f\/inite-dimensional subspace of~$H^1(\partial
M)$.
On the other hand an inf\/inite-dimensional space consisting of coclosed $1$-forms.
Thus we give a~complementary view, although that was not our original aim.
The proof in~\cite{CMR1} is short and brief\/ly describes the main ideas.
We give a~more detailed proof, since our aim is to exhibit the explicit application of an axiomatic system that seems
sketched in~\cite{CMR}.
We give explicit calculations in terms of local coordinate decomposition.
Finally our results extend to regions that are manifolds with corners.
This is essential for the physically most relevant case of gluing,
where the component manifolds as well as the composite manifold have
the topology of a~ball.

\looseness=1
A related work~\cite{AM}, describes Killing vector f\/ield acting on $1$-forms with Dirichlet and Neumann boundary
conditions.
The author thanks the referee for pointing out the reference~\cite{Sa}, where gauge action is described for spin
manifolds with boundary in the context of M theories.

{\bf Description of sections.} Section~\ref{sec:1} consists of a~review of the symplectic formalism for classical f\/ield
theories together with an exposition of the language of abelian gauge f\/ield theories and manifolds with corners.
In Section~\ref{sec:axioms} we exhibit the axioms of an abelian gauge f\/ield theory which is divided in two cases: the
case where regions are considered as smooth manifolds with boundary and the case where regions are manifolds with
corners.
We construct the abelian theory using this axiomatic system.
In Section~\ref{sec:2}, we focus on the kinematics of gauge f\/ields.
This section involves local considerations where Moser's arguments on the transport f\/low for volume forms is used.
A~similar argument due to Dacorogna--Moser for manifolds with boundary is crucial for the corners case.
Dynamics of gauge f\/ields is explored in Section~\ref{sec:Dynamics}.
We describe the symplectic reduction of the space of boundary conditions and emphasize the proofs of the Lagrangian
embedding of solutions.
This last result uses Friedrichs--Morrey--Hodge theory adapted to the case of corners.
Finally we review the special case of Yang--Mills theory in dimension 2 in Section~\ref{sec:4} for illustration.

\section{Classical abelian gauge f\/ield theories}
\label{sec:1}

For the sake of completeness, we summarize the symplectic formalism for Lagrangian f\/ield theories in the following
paragraphs.
Local descriptions for the case of the space of Dirichlet--Neumann conditions appear in~\cite{KT}.
On the other hand the discussion of the space of germs of solutions in the axiomatic setting appears in~\cite{O1, O}.
Parallel developments appear also in~\cite{CMR}.
We adopt an abstract coordinate-free description of the (pre)symplectic structure for boundary data, by means of
a~suitable cohomological point of view.

\subsection{The symplectic setting for classical Lagrangian f\/ield theories}

Classical f\/ield theory assumes that over an~$n$-dimensional space-time region~$M$, there exists a~``conf\/iguration
space'', $K_M$, of f\/ields $\varphi\in K_M$.
The word ``space'' used for referring to $K_M$ usual\-ly denotes an inf\/inite-dimensional Fr\`echet manifold, def\/ined as
a~space of sections of a~smooth bundle~$E$ over~$M$.
It also assumes the existence of a~\emph{Lagrangian density}, $\Lambda\in\Omega^n(J^1M)$, depending on the f\/irst-jet
$j^1\varphi\in J^1M$, i.e., on the f\/irst order derivatives~$\partial\varphi$ and on the values of the f\/ields~$\varphi$.
The \emph{action} corresponding to the Lagrangian density is then def\/ined as
\begin{gather*}
S_M(\varphi)=\int_Mj^1(\varphi)^*\Lambda.
\end{gather*}
On the other hand we consider the factorization of the space of~$k$-forms over the~$l$-jet mani\-fold~$J^lM$ as
\begin{gather*}
\Omega^k\big(J^lM\big)=\bigoplus_{r=0}^k\Omega^r_H\big(J^lM\big)\otimes\Omega^{k-r}_V\big(J^lM\big),
\end{gather*}
where the complex $\Omega^k_H(J^lK_M)$ (resp.\
$\Omega^k_V(J^lK_M)$) corresponds to \emph{horizontal $($resp.\
vertical$)$~$k$-forms}.
For instance, using local coordinates $x^i$, $i=1,\dots, n$, for the manifold~$M$, take $(x^i;u^a;u^a_i)$ as local
coordinates for~$J^1M$.
Then \emph{horizontal forms} have as a~basis the exterior product of the~$dx^i$.
Meanwhile for \emph{vertical forms}  in~$J^1M$,   have as basis the exterior product of~$du^a$,~$du^a_i$.

The \emph{horizontal $($resp.\
vertical$)$ differential} is induced by the coordinate decomposition
\begin{gather*}
d_H\colon \ \Omega^k_H\big(J^lM\big)\rightarrow \Omega^{k+1}_H\big(J^{l+1}M\big)
\end{gather*}
(resp.~$d_V:=d-d_H$).
For instance, for horizontal $0$-forms we have
\begin{gather*}
d_H:=\sum\limits_{i=1}^n\left(\frac{\partial}{\partial x_i}+\sum\limits_{a=1}^ru_i^a\frac{\partial}{\partial
u^a}\right)dx^i\colon \ \Omega^0_H\big(J^0M\big)\rightarrow\Omega^1_H\big(J^1M\big),
\end{gather*}
where~$r$ equals the dimension of each f\/iber of the bundle~$E$.
Thus, vertical~$k$-forms vanish on horizontal vector f\/ields $\vec{X}$ such that $d_V(\vec{X})=0$.
This decomposition yields a~\emph{variational bicomplex}, see for instance~\cite{Ak},
\begin{gather*}
\xymatrix{
	0&&0&\dots\\
	\Omega^n_H(J^1M)\ar[rr]_{d_V}\ar[u]^{d_H} &&
	\Omega^n_H(J^1M)\otimes\Omega^{1}_V(J^2M)\ar[u]^{d_H}\ar[r]& \dots\\
	\Omega^{n-1}_H(J^0M)\ar[rr]^{d_V}\ar[u]^{d_H} &&
	\Omega^{n-1}_H(J^0M)\otimes\Omega^{1}_V(J^1M)\ar[u]^{d_H}\ar[r]& \dots\\
	\vdots\ar[u]^{d_H} &&
	\vdots\ar[u]^{d_H}& \vdots\\
}
\end{gather*}
Denote the \emph{space of Euler--Lagrange solutions}  as
\begin{gather*}
A_M=\left\{\varphi\in K_M \,|\, \big(j^2\varphi\big)^*\left(d_V\Lambda\right)=0\right\}.
\end{gather*}
In the case we are dealing with the space of connections~$A_M$ is an af\/fine space.
The corresponding linear space is denoted as $L_M$.

Consider the image $d_V\Lambda\in\Omega^n_H(J^1M)\otimes\Omega^{1}_V(J^2M)$, of the Lagrangian density,
$\Lambda\in\Omega^n_H(J^1M)$.
Take a~preimage
\begin{gather*}
\theta_\Lambda\in d_H^{-1}\circ d_V\Lambda\in \Omega^{n-1}_H\big(J^0M\big)\otimes\Omega^{1}_V\big(J^1M\big).
\end{gather*}
Of course the representative $\theta_\Lambda\in d_H^{-1}\circ d_V\Lambda$ depends just on the $d_H$-cohomology class of
the Lagrangian density.
By integration by parts, the dif\/ferential of the action~$dS_M$, evaluated on variations $\delta\varphi=X\in T_\varphi
K_M$, may be decomposed as
\begin{gather*}
dS_M(\delta\varphi)=(dS_M)_\varphi ({X} )=\int_M\big(j^2\varphi\big)^*\big(\iota_{(j^2\vec{X})}d_V\Lambda\big)+
\int_{\partial M}\big(j^1\varphi\big)^*\big(\iota_{\vec{X}}\theta_\Lambda\big).
\end{gather*}
Locally each variation $\delta\varphi$ is identif\/ied with a~vector f\/ield, $\vec{X}$, along the section~$j^1\varphi$ in~$J^1M$.
This~$\vec{X}$ in turn induces a~vector f\/ield $j^2\vec{X}$, the $2$-jet prolongation of the vector f\/ield~$\vec{X}$,
along $j^2\varphi$, on the $2$-jet manifold~$J^2M$.
Both~$\vec{X}$ and~$j^2\vec{X}$ vanish on horizontal $1$-forms.
This shows that total variations consist of two contributions.
One kind of variation is the one localized on the bulk of the f\/ields corresponding to Euler--Lagrange equations.
Another kind of contribution to the variation comes from the f\/ield and its normal derivatives on the boundary~$\partial
M$.

Let us concentrate on the boundary term of the variation.
The calculus on the $1$-jet total space, $J^1M$, translates to the calculus on the inf\/inite-dimensional space,~$K_M$, so
that $\theta_\Lambda$ induces a~$1$-form
\begin{gather*}
(dS_M)_\varphi({X})=\int_{\partial M}\big(j^1\varphi\big)^*\big(\iota_{\vec{X}}\theta_\Lambda\big)
\end{gather*}
for variations $X\in T_{\varphi} A_M$ of $1$-jets of solutions restricted to the boundary.
This enables us to consider a~$1$-form $dS_M$, for variations ${X}\in T_\varphi A_M$.

For an $(n-1)$-dimensional boundary manifold~$\Sigma$, the boundary conditions for solutions on a~\emph{tubular
neighborhood} $\Sigma_\varepsilon\cong\Sigma\times [0,\varepsilon]$, can be described as germs of solutions.

The af\/f\/ine \emph{space of germs of solutions on the boundary,} and the corresponding linear space are def\/ined as the
injective limits
\begin{gather*}
\tilde{A}_{\Sigma}:=\varinjlim A_{\Sigma_\varepsilon},
\qquad
\tilde{L}_{\Sigma}:=\varinjlim L_{\Sigma_\varepsilon},
\end{gather*}
where the inclusion of tubular neighborhoods, $\Sigma_\varepsilon\!\!\subset\! \Sigma_{\varepsilon'}$, for $\varepsilon
\!<\!\varepsilon'$, induces an inclusion \mbox{$A_{\Sigma_{\varepsilon'}}\!\!\subset \!A_{\Sigma_{\varepsilon}}$}.
Similarly, there is an inclusion for the linear spaces $L_{\Sigma_{\varepsilon'}}\subset L_{\Sigma_{\varepsilon}}$.

The submersion of variations of germs $\tilde{X}\in T_\varphi\tilde{A}_{\Sigma}$, onto variations of jets $\vec{X}\in
T_{\varphi}A_{\Sigma}$, leads to the def\/inition of the $1$-form on $\tilde{A}_{\partial M}$,
\begin{gather*}
\big(\tilde{\theta}_{\Sigma}\big)_\varphi(\tilde{X}):=(dS_M)_\varphi({X}).
\end{gather*}
Ultimately, our purpose is to consider the \emph{presymplectic} structure on $\tilde{A}_{\Sigma}$,
\begin{gather*}
\tilde{\omega}_{\Sigma}=d\tilde{\theta}_{\Sigma}.
\end{gather*}
There are degeneracies of the presymplectic structure $\tilde{\omega}_{\Sigma}$ due to the degeneracy of the Lagrangian
density and the degeneracies arising from considering arbitrary order derivatives for the germs of solutions.
We suppose that these degeneracies altogether can be eliminated by taking the quotient by $K_{\omega_\Sigma}:=\ker
{\omega_\Sigma}$.
Then we obtain a~symplectic space $(A_{\Sigma}, \omega_{\Sigma})$.

Consider an action map $S_M(\varphi)$ def\/ined for connections~$\varphi$ of a~principal bundle~$P$ over~$M$ with compact
abelian structure group~$G$.
We denote as $A_M$, the space of Euler--Lagrange solutions in the interior of the region~$M$.
In general, we suppose that $\partial M$ is not empty.
Hence when we restrict the action functional~$S_M$, from the conf\/iguration f\/ield space $K_M$ to the space of solutions
$A_M$, it induces a~non-constant map
\begin{gather*}
S_M\colon \ A_M\rightarrow\mathbb{R}.
\end{gather*}
On the other hand we have the groups, $G_M$, of \emph{gauge symmetries on regions} acting in $A_M$ the solutions on the
bulk that come from the Euler--Lagrange variational symmetries of the Lagrangian density, see~\cite[Def\/inition~2.3.1]{Ak}.
Inf\/initesimal symmetries can be identif\/ied with~$G$-invariant vertical vector f\/ields on~$P$, i.e., with vertical vector
f\/ields acting on~$J^1P/G$ and preserving the Lagrangian density.

By taking the tubular neighborhood, $\Sigma_\varepsilon$ as the region~$M$, those symmetries by the group~$G_{\Sigma_\varepsilon}$ act on germs of solutions in~$A_{\Sigma_\varepsilon}$ hence in~$\tilde{A}_{\Sigma}$.
By taking the quotient by the stabilizer of the~$\tilde{A}_{\Sigma}$, we obtain a~group of \emph{gauge symmetries on
hypersurfaces},
\begin{gather*}
\tilde{G}_{\Sigma}:=\varinjlim G_{\Sigma_\varepsilon}
\end{gather*}
acting on $\tilde{A}_{\Sigma}$.

Once we have taken the quotient of the space of germs~$\tilde{A}_{\Sigma}$,  and its corresponding linear space~$\tilde{L}_{\Sigma}$, by the degeneracy space~$K_{\omega_\Sigma}$, we get a~space~$A_{\Sigma}$, and a~gauge group~$G_{\Sigma}$ acting on~$A_{\Sigma}$.
The group $\tilde{G}_{\Sigma}$ decomposes into two kind of symmetries: those coming from the degeneracy of the
presymplectic structure and those preserving the symplectic structure coming from vector f\/ields preserving the
Lagrangian density.
This means that there is a~normal subgroup~$K_{\omega_\Sigma}\subset \tilde{G}_{\Sigma}$, that takes into account all
degeneracies.
The $K_{\omega_\Sigma}$-orbits on~$\tilde{A}_\Sigma$ consist of the integral leafs of the characteristic distribution
generated by the kernel of the presymplectic structure~$\tilde{\omega}_\Sigma$.
Meanwhile, the quotient group~$G_\Sigma$ acts by symplectomorphisms on~$A_\Sigma$ with respect to the symplectic
structure~$\omega_\Sigma$.

\subsection{Regions with and without corners}

In the following presentation of the axiomatic system for classical Lagrangian f\/ield theories, we will consider regions
and hypersurfaces as \emph{manifolds with corners}.
We adopt the def\/inition of stratif\/ied spaces in~\cite{AGO}.
In order to establish the notation that will be used along this work we sketch the def\/initions of manifolds with
corners, for a~more detailed description see the cited reference.

\emph{Hypersurfaces} are $(n-1)$-dimensional topological manifolds~$\Sigma$ which decompose as a~union of
$(n-1)$-dimensional manifolds with corners,
\begin{gather*}
\Sigma=\cup_{i=1}^m\Sigma^i\cong\sqcup_{i=1}^m\check{\Sigma}^i/\sim_{\mathcal{P}}.
\end{gather*}
This union in turn is obtained by gluing of $(n-1)$-dimensional manifolds with corners:
$\check{\Sigma}^i$, $\check{\Sigma}^j$, along pairs of $(n-2)$-faces.
This can be done by means of an equivalence relation $\sim_{\mathcal{P}}$, def\/ined by a~certain set $\mathcal{P}$ of
pairs $(i,j)$, $i\neq j$.
More precisely, non trivial equivalence identif\/ications take place for the set
\begin{gather*}
\cup_{(i,j)\in \mathcal{P}}\Sigma^{ij}:=\cup_{(i,j)\in \mathcal{P}}\Sigma^i\cap\Sigma^j.
\end{gather*}
This means that gluings of the faces ${\Sigma}^i$, ${\Sigma}^j$, take place at $(n-2)$-faces ${\Sigma}^{ij}\subset
\partial{\Sigma}^i$, ${\Sigma}^{ji}\subset\partial{\Sigma}^j$, ${\Sigma}^{ji}\cong {\Sigma}^{ij}$.

Consider a~hypersurface~$\Sigma$ as a~stratif\/ied space consisting of a~union $\cup_{i=1}^m\Sigma^i$ of manifolds with
corners $\check{\Sigma^i}$ identif\/ied along their their faces $\partial\check{\Sigma^i}$.
Denote the structure of \emph{stratified spaces}, as $|\Sigma|$ respectively.
For a~stratif\/ied space $|\Sigma|$ we denote the~$k$-dimensional \emph{skeleton} as $|\Sigma|^{(k)}$,
$k=0,1,2,\dots,n-1$,
notice that $|\Sigma|^{(n-1)}\cong\Sigma$ and
\begin{gather*}
|\Sigma|^{(n-2)}=\cup_{(i,j)\in\mathcal{P}}\Sigma^{ij}\subset \Sigma
\end{gather*}
corresponds to the corners set.
We adopt the notation for the set of~$k$-dimensional \emph{faces} as $|\Sigma|^{k}$.
Thus
\begin{gather*}
|\Sigma|^{n-1}=\left\{\check{\Sigma}^1,\dots,\check{\Sigma}^m\right\}
\end{gather*}
is the set of $(n-1)$-dimensional faces and
\begin{gather*}
|\Sigma|^{n-2}=\left\{\check{\Sigma}^{ij} \,|\, (i,j)\in \mathcal{P}\right\}
\end{gather*}
is the set of $(n-2)$-faces.
Here $\check{\Sigma}^{ij}\subset\check{\Sigma}^i$
is the preimage of the corner $\Sigma^{ij}=\Sigma^i\cap\Sigma^j\subset \Sigma$, $(i,j)\in\mathcal{P}$.

A \emph{region} is an~$n$-dimensional manifold with corners~$M$.
Its boundary $\partial M$, is a~topological manifold.
The corresponding stratif\/ied space structures are $|M|$, $|\partial M|$.
Each hypersurface $\Sigma\subset \partial M$ consists of the union of faces $\Sigma^i\subset \partial M$, which are manifolds with
corners.

An abstract closed smooth hypersurface~$\Sigma$, not necessarily related to a~region~$M$, may be considered as
a~component of the boundary of a~cylinder $\Sigma\times[0,\varepsilon]$, $\partial \Sigma=\varnothing$~\cite{Mi}.

The notion of a~cylinder can be generalized for a~manifold with corners~$\Sigma$, $\partial \Sigma\neq\varnothing$.
A~\emph{regular cylinder} consists of
\begin{gather}
\label{eqn:regcyl}
\widehat{\Sigma}_\varepsilon:=\big\{(s,t)\in\Sigma\times[0,\varepsilon] \,|\, t\in[0,\epsilon(s)\varepsilon],
s\in\Sigma\big\}\subset\Sigma\times[0,\varepsilon],
\end{gather}
where $\epsilon\colon \Sigma\rightarrow[0,1]$ is an increasing smooth function such that $\epsilon^{-1}(0)=\partial \Sigma$
and $\Sigma^{\epsilon}:=\epsilon^{-1}(1)\subset\Sigma$ is a~smooth retract deformation of~$\Sigma$.

Thus~$\Sigma$ corresponds to one \emph{face} of the~$n$-dimensional manifold with corners given by the regular cylinder
$\widehat{\Sigma}_\varepsilon$.
In general $\partial \Sigma$, $\partial\Sigma^i$ may be nonempty.

For smooth hypersurfaces $\Sigma\subset \partial M$ we consider tubular neighborhoods~\cite{Mi},
$\Sigma_\varepsilon\subset M$ with dif\/feomorphisms
\begin{gather*}
X\colon \ \Sigma\times[0,\varepsilon]\rightarrow \Sigma_\varepsilon.
\end{gather*}
On the other hand, a~\emph{regular tubular neighborhood} for a~face $\Sigma\subset\partial M$, consists of
a~homeomorphism that becomes a~dif\/feomorphism outside the corners $\partial\Sigma\subset\widehat{\Sigma}_\varepsilon$
\begin{gather*}
X\colon \ \widehat{\Sigma}_\varepsilon\rightarrow \Sigma_\varepsilon
\end{gather*}

Recall that the \emph{corners} of the region~$M$ lie in the union of the $(n-2)$-dimensional submanifolds,
$\cup_{(i,j)\in \mathcal{P}}\Sigma^{ij}$.

The \emph{gluing} of a~region~$M$ along two nonintersecting faces $\Sigma_0$, $\Sigma_0'$, can be def\/ined.
The more general gluing along two nonintersecting \emph{hypersurfaces} $\Sigma$, $\Sigma'$, may also be def\/ined.
Nonetheless, when we consider, for instance, the gluing of Riemannian metrics, this gluing along general hypersurfaces
may be problematic.
For if we glue faces with non intersecting boundaries $\partial \Sigma_0\cap\partial\Sigma_0'=\varnothing$, then conic
singularities of the metric along the corners may arise in the resulting space-time region.

{\it We consider gluings along nonintersecting faces and do not consider gluings along hypersurfaces.}

\section{Axiomatic system proposal}
\label{sec:axioms}

Now we give a~detailed description of the axiomatic framework for classical gauge f\/ield theories.
Axioms~\ref{ax:1}--\ref{ax:10} describe the kinematics of the classical theory, while Axioms~\ref{ax:9}--\ref{ax:12}
describe the dynamics for gauge f\/ields.

\subsection{GBF Axioms}

We consider space-time regions~$M$ that are manifolds with corners of dimension~$n$, as well as hypersurfaces~$\Sigma$
that are topological $(n-1)$-dimensional topological manifolds with stratif\/ied space structure~$|\Sigma|$.

\begin{enumerate}[label=A\arabic*]\itemsep=0pt

\item
\label{ax:1}
\emph{Affine structure.} For space-time regions~$M$ we have the af\/f\/ine spaces $A_M$ with the asso\-cia\-ted linear spaces
${L}_M$ of Euler--Lagrange solutions.
On the other hand, for hypersurfaces~$\Sigma$ we have af\/f\/ine spaces $A_\Sigma$ with associated linear spaces~$\tilde{L}_\Sigma$, of bounda\-ry conditions.
There are also af\/f\/ine maps $\tilde{a}_M\colon {A}_M\rightarrow \tilde{A}_{\partial M}$, as well as linear maps
$\tilde{r}_M\colon {L}_M\rightarrow \tilde{L}_{\partial M}$.

\item
\label{ax:2}
\emph{Presymplectic structure.} For every hypersurface $\Sigma\subset \partial M$, there is a~presymplectic structure
$\tilde{\omega}_\Sigma$ on $\tilde{A}_\Sigma$ invariant under $\tilde{L}_\Sigma$ actions.
Equivalently we can consider $\tilde{L}_\Sigma$ as a~presymplectic vector space with presymplectic structure denoted also as
$\tilde{\omega}_\Sigma$.

\item
\label{ax:3}
\emph{Symplectic structure.} There is a~group $K_{{\omega}_{\Sigma}}$ acting freely by translations on
$\tilde{A}_\Sigma$, such that~$K_{{\omega}_{\Sigma}}$ is isomorphic to the~closed linear subspace
$\ker\tilde{\omega}_\Sigma\subset \tilde{L}_\Sigma$.
So $\tilde{\omega}_\Sigma$ induces a~{symplectic} structure,~$\omega_\Sigma$, on the orbit space
\begin{gather*}
{A}_\Sigma:=\tilde{A}_\Sigma/K_{{\omega}_{\Sigma}}.
\end{gather*}
This space is an af\/f\/ine space modeled on the linear space ${L_\Sigma}:=\tilde{L}_\Sigma/K_{\omega_\Sigma}$.
By taking the quotients, the maps $\tilde{a}_M$ and $\tilde{r}_M$ induce af\/f\/ine and linear maps $a_M\colon A_M\rightarrow A_{\partial
M}$, $r_M\colon A_M\rightarrow A_{\partial M}$, respectively.

\item
\label{ax:4}
\emph{Symplectic potential.} There is a~symplectic potential, i.e., an ${L_\Sigma}$-valued $1$-form
$\theta_\Sigma(\varphi,\cdot)$ for each $\varphi\in {A_\Sigma}$, identif\/ied with a~linear map
$\theta_\Sigma(\varphi,\cdot)\colon {L_\Sigma}\rightarrow \mathbb{R}$.
There is also a~bilinear map $[\cdot,\cdot]_\Sigma\colon {L_\Sigma}\times {L_\Sigma}\rightarrow \mathbb{R}$ such that
\begin{gather*}
[\phi,\phi']_\Sigma+\theta_\Sigma(\eta,\phi')=\theta_\Sigma(\phi+\eta,\phi'),
\qquad
\eta\in {A_\Sigma}, \phi,\phi'\in {L_\Sigma}
\end{gather*}
and
\begin{gather}
\label{eqn:2}
\omega_\Sigma(\phi,\phi')=\frac{1}{2}[\phi,\phi']_\Sigma-\frac{1}{2}[\phi',\phi]_\Sigma,
\qquad
\phi,\phi'\in {L_\Sigma}.
\end{gather}
There exists an action map $S_M\colon A_M\rightarrow \mathbb{R}$, such that{\samepage
\begin{gather}
\label{eqn:00}
S_M(\eta)=S_M(\eta')-\frac{1}{2}\theta_{\partial M}(\eta,\eta -\eta')-\frac{1}{2}\theta_{\partial M}(\eta',\eta-\eta')
\end{gather}
and also $S_M(\eta)=S_M(\eta')$ for $a_M(\eta)=a_M(\eta')$.}

\item
\label{ax:5}
\emph{Involution.} For each hypersurface~$\Sigma$ there exists an involution $A_\Sigma\rightarrow
A_{\overline{\Sigma}}$, where $\overline{\Sigma}$ is the hypersurface with reversed orientation.
There is also a~linear involution $L_\Sigma\rightarrow L_{\overline{\Sigma}}$.
We have: $\theta_{\overline{\Sigma}}(\eta,\phi)=-\theta_\Sigma(\eta,\phi)$ and
$[\phi,\phi']_{\overline{\Sigma}}=-[\phi,\phi']_\Sigma$.

\item
\label{ax:6}
\emph{Disjoint regions.} For a~disjoint union, $M=M_1\sqcup M_2$, there is a~bijection $A_{M_1}\times A_{M_2}\rightarrow
A_M$ and compatible linear isomorphism $L_{M_1}\times L_{M_2}\rightarrow L_M$, such that $a_M=a_{M_1}\times a_{M_2}$ and
$r_M=r_{M_1}\times r_{M_2}$, satisfy associative conditions.
For the action map we have $S_M=S_{M_1}+S_{M_2}$.

\item
\label{ax:7}
\emph{Factorization of fields on hypersurfaces.} For a~hypersurface~$\Sigma$ obtained as the quotient
$\check{\Sigma}^1\sqcup\dots\sqcup\check{\Sigma}^k$ by an equivalence relation $\sim_{\mathcal{P}}$, def\/ine
$A_{|\Sigma|^{n-1}}:=A_{\check{\Sigma}^1}\times\dots\times A_{\check{\Sigma}^m}$,
$L_{|\Sigma|^{n-1}}:=L_{\check{\Sigma}^1}\oplus\dots\oplus L_{\check{\Sigma}^m}$. Then there are af\/f\/ine gluing maps
$a_{\Sigma,{|\Sigma|^{n-1}}}\colon A_{\Sigma}\rightarrow A_{|\Sigma|^{n-1}}$, and compatible linear maps
$r_{\Sigma,{|\Sigma|^{n-1}}}\colon L_{\Sigma}\rightarrow L_{|\Sigma|^{n-1}}$ with commuting diagrams
\begin{gather*}
	\xymatrix{
		&A_{{|\Sigma|^{n-1}}}\ar[d]\\
		A_{\Sigma}\ar[r]\ar@{-->}[ru]^{a_{\Sigma,|\Sigma|^{n-1}}}& A_{\check{\Sigma}^i}
	}\qquad
	\xymatrix{
		&L_{{|\Sigma|^{n-1}}}\ar[d]\\
		L_{\Sigma}\ar[r]\ar@{-->}[ru]^{r_{\Sigma,|\Sigma|^{n-1}}}& L_{\check{\Sigma}^i}
	}
\end{gather*}
We also have the relation
\begin{gather}
[\cdot,\cdot]_{\Sigma}=r_{\Sigma;|\Sigma|^{n-1}}^*\big([\cdot,\cdot]_{\check{\Sigma}^1}+\dots+[\cdot,\cdot]_{\check{\Sigma}^m}\big),
\label{eqn:[]}
\qquad
\theta_{\Sigma}=r_{\Sigma;|\Sigma|^{n-1}}^*\big(\theta_{\check{\Sigma}^1}+\dots+\theta_{\check{\Sigma}^m}\big)
\end{gather}
we denote $A_{\Sigma^i}$ as the image of $A_\Sigma$ into $A_{\check{\Sigma}^i}$, and similarly $L_{\Sigma^i}$.

\item
\label{ax:8}
\emph{Gauge action.} There are groups $\tilde{G}_\Sigma$ acting on $\tilde{A}_\Sigma$ preserving the af\/f\/ine structure
and the presymplectic structure $\tilde{\omega}_\Sigma$ such that $K_{{\omega}_{\Sigma}}\trianglelefteq
\tilde{G}_\Sigma$.
The quotient group
\begin{gather*}
{G}_\Sigma:=\tilde{G}_\Sigma/K_{{\omega}_{\Sigma}}
\end{gather*}
acts on $A_\Sigma$, preserving the symplectic structure $\omega_\Sigma$.
There is a~group,~$G_M$, of gauge variational symmetries for $S_M$ acting on the space of solutions $A_M$.
There is a~group homomorphism~$h_M\colon G_M\rightarrow G_{\partial M}$. For the map $a_M\colon A_M\rightarrow A_{\partial M}$ the compatibility of gauge group actions is given by the commuting diagram
\begin{gather*}
\xymatrix{
	A_M\times G_M\ar[r]\ar[d]&A_{\partial M}\times G_{\partial M}\ar[d]\\
	A_{M}\ar[r]&A_{\partial M}\\
}
\end{gather*}
There is also a~compatible action on the corresponding linear spaces $r_M\colon L_M\rightarrow L_{\partial M}$
\begin{gather*}
\xymatrix{
	L_M\times G_M\ar[r]\ar[d]&L_{\partial M}\times G_{\partial M}\ar[d]\\
	L_{M}\ar[r]&L_{\partial M}\\
}
\end{gather*}

\item
\label{ax:10}
\emph{Factorization of gauge actions on hypersurfaces.} For the case of regions with corners there is a~homomorphism
$h_{\Sigma;{|\Sigma|^{n-1}}}\colon G_{\Sigma}\rightarrow G_{|\Sigma|^{n-1}}$ from the direct product group
$G_{|\Sigma|^{n-1}}:=G_{\check{\Sigma}^1}\times\dots \times G_{\check{\Sigma}^m}$ onto $G_{\Sigma}$ coming from
homomorphisms
\begin{gather*}
h_{|\Sigma|^{n-1};\check{\Sigma}^i}\colon \ G_{|\Sigma|^{n-1}}\rightarrow G_{\check{\Sigma}^i}
\end{gather*}
and commuting diagrams
\begin{gather*}
\xymatrix{
	&G_{|\Sigma|^{n-1}}\ar[d]\\
	G_{\Sigma}\ar[r]\ar@{-->}[ru]& G_{\check{\Sigma}^i}
}
\end{gather*}
and
\begin{gather*}
\xymatrix{
	&&A_{|\Sigma|^{n-1}}\ar[ddr]&\\
	&&A_{|\Sigma|^{n-1}}\times G_{|\Sigma|^{n-1}}\ar[d]\ar[u]&\\
	A_{\Sigma}\ar[uurr]\ar@/_0.5cm/[rrr]&
	A_{\Sigma}\times G_{\Sigma}\ar[l]\ar[r]\ar[ru]&
	A_{\check{\Sigma}^i}\times G_{\check{\Sigma}^i}\ar[r]&A_{\check{\Sigma}^i}
}
\end{gather*}
and similar commuting diagrams for actions on linear spaces
\begin{gather*}
\xymatrix{
	&&L_{|\Sigma|^{n-1}}\ar[ddr]&\\
	&&L_{|\Sigma|^{n-1}}\times G_{|\Sigma|^{n-1}}\ar[d]\ar[u]&\\
	L_{\Sigma}\ar[uurr]\ar@/_0.5cm/[rrr] &
	L_{\Sigma}\times G_{\Sigma}\ar[l]\ar[r]\ar[ru]&
	L_{\check{\Sigma}^i}\times G_{\check{\Sigma}^i}\ar[r] &  L_{\check{\Sigma}^i}
}.
\end{gather*}
There is an involution of the gauge groups $G_\Sigma\rightarrow G_{\overline{\Sigma}}$, compatible with the action.

We denote the image $h_{|\Sigma|^{n-1};\check{\Sigma}^i} (G_\Sigma )\subset G_{\check{\Sigma}^i}$ as $G_{\Sigma^i}$.

\item
\label{ax:9}
\emph{Lagrangian relation modulo gauge.} Let ${A}_{\tilde{M}}$ be the image ${a}_M({A}_M)\subset {{A}}_{\partial M}$ of
boundary conditions on $\partial M$ that extend to solutions on the bulk~$M$.
Let ${L}_{\tilde{M}}$ be the corresponding linear subspace that is the image ${r}_M({L}_M)\subset {{L}}_{\partial M}$.
The subspace ${L}_{\tilde{M}}\subset L_{\partial M}$ is Lagrangian.

The zero component of the $G_{\partial M}$-orbit is isomorphic to $C_{\partial M}^\bot$, the symplectic orthogonal
complement of a~coisotropic subspace $C_{\partial M}\subset L_{\partial M}$.
There is a~Lagrangian reduced subspace isomorphic to
\begin{gather*}
{L}_{\tilde{M}}\cap C_{\partial M}/{L}_{\tilde{M}}\cap C_{\partial M}^\bot
\end{gather*}
of the symplectic reduced space $C_{\partial M}/C_{\partial M}^\bot$.

\item
\label{ax:11}
\emph{Locality of gauge fields.} Let $M_1$ be the region that can be obtained by the gluing of~$M$ along the disjoint
faces, $\Sigma_0,\overline{\Sigma_0'}\subset \partial M$, where $\Sigma_0'\cong\Sigma_0$.
Then there is an injective af\/f\/ine map, $a_{M;\Sigma_0,\overline{\Sigma_0'}}\colon A_{M_1}\hookrightarrow A_M$, a~compatible
linear map, $r_{M;\Sigma_0,\overline{\Sigma_0'}}\colon L_{M_1}\hookrightarrow L_M$, and a~homomorphism
$h_{M;\Sigma_0,\overline{\Sigma_0'}}\colon G_{M_1}\hookrightarrow G_M$, with exact sequences
\begin{gather*}
A_{M_1}\hookrightarrow A_M \rightrightarrows A_{\Sigma_0},
\qquad
L_{M_1}\hookrightarrow L_M \rightrightarrows L_{\Sigma_0},
\qquad
G_{M_1}\hookrightarrow G_M \rightrightarrows G_{\Sigma_0},
\end{gather*}
where we consider the involution $A_{\overline{\Sigma'_0}}\rightarrow A_{\Sigma_0}$, for the second arrow on the double
map.
Recall that $A_{\Sigma_0}$ is the image in $A_{\check{\Sigma}_0}$.
We consider the gluing of the actions
\begin{gather*}
\xymatrix{
	A_{M_1}\times G_{M_1}\ar[r]\ar[d]&A_M\times G_M\ar@/^/[r]\ar@/_/[r]\ar[d]&A_{\Sigma_0}\times G_{\Sigma_0}\ar[d]\\
	A_{M_1}\ar[r]&A_M\ar@/^/[r]\ar@/_/[r]&A_{\Sigma_0}\\
}
\end{gather*}
compatible with the actions on linear spaces
\begin{gather*}
\xymatrix{
	L_{M_1}\times G_{M_1}\ar[r]\ar[d]&L_M\times G_M\ar@/^/[r]\ar@/_/[r]\ar[d]&L_{\Sigma_0}\times G_{\Sigma_0}\ar[d]\\
	L_{M_1}\ar[r]&L_M\ar@/^/[r]\ar@/_/[r]&L_{\Sigma_0}\\
}
\end{gather*}
and also $S_{M_1}=S_M \circ a_{M;\Sigma_0;\overline{\Sigma_0'}}$.

\item
\label{ax:12}
\emph{Gluing of gauge fields.} Let $M_1$, $M$ be regions with corners $M_1$ is obtained by gluing~$M$ along
hypersurfaces $\Sigma_0,\overline{\Sigma'_0}\subset \partial M$.
The following diagrams commute
\begin{gather*}
\xymatrix{
	 A_{M_1} \ar[r]\ar[d]& A_{M}\ar[d]\\
	A_{\partial M_1}\ar[d]& A_{\partial M}\ar[d]\\
	 A_{|\partial M_1|^{n-1}} & A_{|\partial M|^{n-1}}\ar[l]
 }\quad
 \xymatrix{
 	G_{M_1} \ar[r]\ar[d]& G_M\ar[d]\\
	G_{\partial M_1}\ar[d]& G_{\partial M}\ar[d]\\
 	G_{|\partial M_1|^{n-1}} & G_{|\partial M|^{n-1}}\ar[l]
 }
\quad
\xymatrix{
 L_{M_1} \ar[r]\ar[d]& L_M\ar[d]\\
L_{\partial M_1}\ar[d]& L_{\partial M}\ar[d]\\
 L_{|\partial M_1|^{n-1}} & L_{|\partial M|^{n-1}}\ar[l]
 }
\end{gather*}
where if $|\partial
M|^{n-1}=\check{\Sigma}_0\sqcup\overline{\check{\Sigma'}_0}\sqcup(\check{\Sigma}^1\sqcup\dots\sqcup\check{\Sigma}^r)$,
and $|\partial M_1|^{n-1}=\check{\Sigma}^1\sqcup\dots\sqcup\check{\Sigma}^r$, then the map $a_{|\partial
M|^{n-1},|\partial M_1|^{n-1}}\colon A_{|\partial M|^{n-1}}\rightarrow A_{|\partial M_1|^{n-1}}$ equals the canonical
inclusion
\begin{gather*}
A_{\check{\Sigma}^1}\times\dots \times A_{\check{\Sigma}^r}\subset A_{\check{\Sigma}_0}\times
A_{\overline{\check{\Sigma}'_0}}\times \big(A_{\check{\Sigma}^1}\times\dots \times A_{\check{\Sigma}^r}\big).
\end{gather*}
We have similar inclusions
\begin{gather*}
r_{|\partial M|^{n-1},|\partial M_1|^{n-1}}\colon \ L_{|\partial M|^{n-1}}\rightarrow L_{|\partial M_1|^{n-1}},
\\
h_{|\partial M|^{n-1},|\partial M_1|^{n-1}}\colon \ G_{|\partial M|^{n-1}}\rightarrow G_{|\partial M_1|^{n-1}}.
\end{gather*}
Compatibility for the gluing of the actions of the gauge groups is described by the commuting diagrams:
\begin{gather*}
\xymatrix{
	A_{M_1}\ar[rrr]\ar[dd]&&& A_{M}\ar[dd]
	\\
& A_{\partial M_1}\times G_{M_1} \ar[r]\ar[d]\ar[lu]& A_M\times G_M\ar[d]\ar[ru]&
\\
A_{\partial M_1}\ar[dd]&
A_{\partial M_1}\times G_{\partial M_1}\ar[d]\ar[l]&
A_{\partial M}\times G_{\partial M}\ar[d]\ar[r]&
A_{\partial M}\ar[dd]
\\
&A_{|\partial M_1|^{n-1}}\times G_{|\partial M_1|^{n-1}}\ar[ld] &
A_{|\partial M|^{n-1}}\times G_{|\partial M|^{n-1}}\ar[l]\ar[rd]&
\\
A_{|\partial M_1|^{n-1}}&&&A_{|\partial M|^{n-1}}\ar[lll]
}
\\
\xymatrix{
L_{M_1}\ar[rrr]\ar[dd]&&& L_{M}\ar[dd]
\\
& L_{M_1}\times G_{M_1} \ar[r]\ar[d]\ar[lu]& L_M\times G_M\ar[d]\ar[ru]&
\\
L_{\partial M_1}\ar[dd]&
L_{\partial M_1}\times G_{\partial M_1}\ar[d]\ar[l]&
L_{\partial M}\times G_{\partial M}\ar[d]\ar[r]&
L_{\partial M}\ar[dd]
\\
&L_{|\partial M_1|^{n-1}}\times G_{|\partial M_1|^{n-1}}\ar[ld] &
L_{|\partial M|^{n-1}}\times G_{|\partial M|^{n-1}}\ar[l]\ar[rd]&
\\
L_{|\partial M_1|^{n-1}}&&&L_{|\partial M|^{n-1}}\ar[lll]
}
\end{gather*}
\end{enumerate}

\subsection{Further discussion of the axioms}
\label{subsec:locality}

Axioms~\ref{ax:1}--\ref{ax:7} are just a~restatement of Axioms~C1 to~C6 for a~classical setting of af\/f\/ine
(linear) f\/ield theories in~\cite{O1}.
Some clarif\/ications are added: In Axiom~\ref{ax:2} we consider presymplectic spaces of connections instead of symplectic
spaces.
We do not consider Hilbert space structures since we are not introducing yet prequantization.

Some comments can be made about postulate Axiom~\ref{ax:4}.
The translation rule of the $1$-form $\theta_{\partial M}$ can be deduced from the translation rule for the dif\/ferential
$dS_M$ of the action map.
This in turn can be deduced from~\eqref{eqn:00}.
This last relation could be stated as a~primordial property and arises from considering a~quadratic Lagrangian
density~$\Lambda$.
The af\/f\/ine structure for the space of solutions~$A_M$ can also be deduced from this condition on~$\Lambda$.

In Axiom~\ref{ax:7} we adapt the decomposition stated in Axiom~C3 for the corners case.

The set of corners correspond to the $(n-2)$-dimensional faces $\Sigma^{ij}:=\Sigma^i\cap\Sigma^j$,
$(i,j)\in\mathcal{P}$.
The lack of surjectivity for dotted arrows in Axiom~\ref{ax:7} comes from the non dif\/ferentiability of the
hypersurface~$\Sigma$ along the corners $|\Sigma|^{(n-2)}$ in the intersections $\Sigma^i\cap\Sigma^j$, $(i,j)\in\mathcal{P}$.

Axiom~\ref{ax:8} introduces the gauge symmetries.
Axiom~\ref{ax:10} presents the decomposition and involution properties for gauge actions on the boundary.
Finally, Axioms~\ref{ax:11} and~\ref{ax:12} are derivations for the locality and gluing rule of gauge f\/ields arising
from the gluing Axiom~C7.

\emph{Locality} arguments for gauge f\/ields is used in Axioms~\ref{ax:8},~\ref{ax:11} and~\ref{ax:12}.
They deserve further clarif\/ication.
For instance in Axiom~\ref{ax:11}, the existence of the exact sequence is not trivial and it is derived from locality for
connections in~$A_M$ and gauge actions in~$G_M$.
From the inclusions $\partial M_\varepsilon\subset M$ of regular tubular neighborhoods we get the following exact
sequences
\begin{gather*}
\xymatrix{
&&A_{\Sigma_\varepsilon}\\
A_{M_1-(\Sigma_\varepsilon\cup\Sigma'_\varepsilon)}\ar@{^{(}->}[r]&
A_M\ar@/^/@{->>}[ru]\ar@/_/@{->>}[rd]&\\
&&A_{\Sigma_\varepsilon'}\ar@{=}[uu]
}
\end{gather*}
If we consider the maps
\begin{gather*}
\xymatrix{
A_{\Sigma_\varepsilon}\ar@{^{(}->}[r]&\tilde{A}_\Sigma \ar@{->>}[r] &A_\Sigma,
}
\end{gather*}
then we can induce the sequence proposed in the axiom, when $\varepsilon\rightarrow 0$.
Recall that $\tilde{A}_\Sigma$ is an inductive limit, and $A_\Sigma$ is a~quotient of $\tilde{A}_\Sigma$.
For Axiom~\ref{ax:8} similar arguments using the following commutative diagrams
\begin{gather*}
\xymatrix{
A_M\times G_M\ar[r]\ar[d]&A_{\partial M_\varepsilon}\times G_{\partial M_\varepsilon}\ar[d]\\
A_M\ar[r]&A_{\partial M_\varepsilon}
}
\end{gather*}

Axiom~\ref{ax:8} arises from \emph{locality}: there is an embedding of gauge symmetries in~$M$ as local gauge symmetries
in a~tubular neighborhood $\partial M_\varepsilon$, and then there is an inclusion $G_{\partial M_\varepsilon}\subset
\tilde{G}_{\partial M}$.
Finally symmetries from $G_M$ acting on germs yield symmetries in the quotient group $G_{\partial M}$.

Axiom~\ref{ax:9} encodes the dynamics of gauge f\/ields since it is an adapted version of the Lag\-rangian embedding to the
symplectic space $A_{\partial M}$ considered in Axiom~C5.
In Axiom~\ref{ax:9} we use the notion of reduced Lagrangian space, see~\cite{We}.
We could also postulate this dynamics axiom as follows.

There exists a~symplectic closed subspace $\Phi_{A_{\partial M}}\subset L_{\partial M}$, such that $L_{\tilde{M}}$
intersects transversally the space $\Phi_{A_{\partial M}}$.
Hence $L_{\tilde{M}}\cap \Phi_{A_{\partial M}}\subset \Phi_{A_{\partial M}}$ is a~Lagrangian subspace.
Furthermore every $G_{\partial M}$-orbit intersects $\Phi_{A_{\partial M}}$ is a~discrete set.
We also call $\Phi_{A_{\partial M}}$ a~\emph{gauge-fixing space} for the gauge symmetries $G_{\partial M}$.

\subsection{Simplif\/ications in the absence of corners}

As we mentioned previously for some axioms, namely Axioms~\ref{ax:7},~\ref{ax:10} and~\ref{ax:12}, we will consider separately
two cases:

{\it Smooth case.} Regions~$M$ and hypersurfaces are smooth manifolds of dimension~$n$ and $n-1$ respectively,~$\Sigma$
is closed.

{\it Corners case.} Regions~$M$ are~$n$-dimensional manifolds with corners, and hypersurfaces are $(n-1)$-dimensional
topological manifolds~$\Sigma$ with stratif\/ied space structure~$|\Sigma|$.

We write down explicitly these axioms in the smooth case, where regions~$M$ and hypersurfaces~$\Sigma$ are smooth
manifolds.
\begin{itemize}\itemsep=0pt

\item[A7$'$] Suppose that an $(n-1)$-dimensional hypersurface~$\Sigma$ decomposes as a~disjoint union
\begin{gather*}
\Sigma:=\Sigma^1\sqcup\dots\sqcup\Sigma^m
\end{gather*}
of connected components $\Sigma^1,\dots,\Sigma^m$.
Def\/ine $A_{|\Sigma|^{n-1}}:=A_{\Sigma^1}\times\dots\times A_{\Sigma^m}$, $L_{|
\Sigma|^{n-1}}:=L_{\Sigma^1}\oplus\dots\oplus L_{\Sigma^m}$. Then there are linear and af\/f\/ine isomorphisms respectively
\begin{gather*}
r_{\Sigma;|\Sigma|^{n-1}}\colon \ L_{\Sigma}\rightarrow L_{|\Sigma|^{n-1}},
\qquad
a_{\Sigma;|\Sigma|^{n-1}}\colon \ A_{\Sigma}\rightarrow A_{|\Sigma|^{n-1}}
\end{gather*}
such that~\eqref{eqn:[]} holds.

\item[A9$'$] For the case without corners ${|\Sigma|^{n-1}}\cong\Sigma$ and the direct product group
$G_{|\Sigma|^{n-1}}:=G_{\Sigma^1}\times\dots \times G_{\Sigma^m}$ is isomorphic to $G_{\Sigma}$ with a~gluing
homomorphisms $h_{\Sigma;{|\Sigma|^{n-1}}}\colon G_{\Sigma}\rightarrow G_{|\Sigma|^{n-1}}$ with compatibility commuting
diagrams
\begin{gather*}
\xymatrix{
A_{\Sigma}\times G_{\Sigma}\ar@{<->}[r]\ar[d]&
A_{|\Sigma|^{n-1}}\times G_{|\Sigma|^{n-1}}\ar[d]\\
A_{\Sigma}\ar@{<->}[r]&
A_{|\Sigma|^{n-1}}
}
\end{gather*}
and analogous compatibility diagrams for actions on linear spaces $L_{\partial M}$, $L_{|\partial M|^{n-1}}$.

\item[A12$'$] Let $M_1$,~$M$ be regions without corners as above with gluing along hypersurfaces
$\Sigma$, $\overline{\Sigma'}\subset \partial M$
\begin{gather*}
\xymatrix{
 A_{M_1} \ar[r]\ar[d]& A_{M}\ar[d]\\
A_{\partial M_1}& A_{\partial M}\ar[l]
}\qquad
\xymatrix{
 G_{M_1} \ar[r]\ar[d]& G_M\ar[d]\\
G_{\partial M_1}& G_{\partial M}\ar[l]
 }\qquad
 \xymatrix{
 L_{M_1} \ar[r]\ar[d]& L_M\ar[d]\\
L_{\partial M_1}& L_{\partial M}\ar[l]
 }
\end{gather*}
There is also a~compatibility for the gluing of the actions of the gauge groups
\begin{gather*}
\xymatrix{
A_{M_1}\ar[rrr]\ar[ddd]&&& A_{M}\ar[ddd]
\\
& A_{M_1}\times G_{M_1} \ar[r]\ar[d]\ar[lu]& A_M\times G_M\ar[d]\ar[ru]&
\\
&A_{\partial M_1}\times G_{\partial M_1}\ar[ld] &
A_{\partial M}\times G_{\partial M}\ar[l]\ar[rd]&
\\
A_{\partial M_1}&&&A_{\partial M}\ar[lll]
}
\\
\xymatrix{
 L_{M_1}\ar[rrr]\ar[ddd]&&& L_{M}\ar[ddd]
\\
& L_{M_1}\times G_{M_1} \ar[r]\ar[d]\ar[lu]& L_M\times G_M\ar[d]\ar[ru]&
\\
&L_{\partial M_1}\times G_{\partial M_1}\ar[ld] &
L_{\partial M}\times G_{\partial M}\ar[l]\ar[rd]&
\\
L_{\partial M_1}&&&A_{\partial M}\ar[lll]
}
\end{gather*}
\end{itemize}

\section{Kinematics of gauge f\/ields}
\label{sec:2}

In this section we consider af\/f\/ine f\/ield theories (comment of Axiom~\ref{ax:4}).
The action that is used as a~test case comes from the Lagrangian density.
We consider gauge principal bundles on a~compact manifold~$M$ provided with a~Riemannian metric~$h$, nonempty boundary
$\partial M$ and compact \emph{abelian} f\/iber group~$G$.
We suppose that regions~$M$ are manifolds of dimension $n=\dim M\geq 2$, provided with a~trivial principal bundle~$P$
with abelian structure group $G=U(1)$.

\subsection{Classical abelian action}

Along this subsection we assume the following two descriptions of a~\emph{face}~$\Sigma$ of~{hypersurface}.
\begin{enumerate}\itemsep=0pt
\item[A.] \emph{Smooth case:}~$\Sigma$ is a~smooth closed $(n-1)$-dimensional manifold, or
\item[B.] \emph{Corners case:}~$\Sigma$ is an $(n-1)$-dimensional topological manifold with corners.
\end{enumerate}

Since the bundle is trivial, the space of connections $A_M$ has a~linear structure and can be identif\/ied with $L_M$.
We consider the action
\begin{gather*}
S_M(\varphi)=\int_Md\varphi\wedge\star d\varphi,
\end{gather*}
where $\varphi\in A_M$ is a~connection that is a~solution of the Euler--Lagrange equations in the bulk, i.e., $d^\star
d\varphi=0$.
The corresponding linear space is
\begin{gather*}
L_M = \big\{\varphi\in\Omega^1(M) \,|\, d^\star d\varphi=0 \big\}.
\end{gather*}
Here $\Omega^1(M,\mathfrak{g})\simeq\Omega^1(M)$ denotes $\mathfrak{g}$-valued $1$-forms on~$M$.
These objects fulf\/ill Axiom~\ref{ax:1}.

The identity component of gauge symmetries can be identif\/ied with certain $f\in\Omega^0(M)$ acting~by
$\varphi\mapsto\varphi+df$, thus $G_M^0\cong\Omega^0(M)/\mathbb{R}^{b_0}$, where $\mathbb{R}^{b_0}$ denote the locally
constant functions on~$M$.
We consider hypersurfaces as closed submanifolds $\Sigma\subset \partial M$.
Since $G_M^0$ preserves the action on $A_M$ the requirement mentioned in Axiom~\ref{ax:8} is satisf\/ied.

We will describe an embedding, that in the smooth case is
\begin{gather*}
X\colon \ \Sigma\times [0,\varepsilon]\rightarrow \Sigma_\varepsilon.
\end{gather*}
We also consider a~normal vector f\/ield ${\partial_\tau}$ on $\Sigma_\varepsilon$, whose f\/low lines are the trajectories
$X(\cdot,\tau)\in\Sigma_\varepsilon$, $0\leq\tau\leq\varepsilon$, that are normal to the boundary.
This embedding arises from the solution of the volume preserving evolution problem on~$\Sigma$, solved by Moser's trick,
see~\cite{Mo}.

\begin{lma}
\label{lma:embeddingndim}
{\rm A.~Smooth case.} Let~$\Sigma$ be a~compact closed $(n-1)$-manifold that is a~component of the boundary of
a~Riemannian manifold $\Sigma_\varepsilon$ diffeomorphic to a~cylinder $\Sigma\times[0,\varepsilon]$, provided with
a~Riemannian metric~$h$.
Then there exists an embedding $X\colon \Sigma\times[0,\varepsilon]\rightarrow \Sigma_\varepsilon$ such that:

\begin{enumerate}\itemsep=0pt
\item[$1.$]
The vector field $\partial_\tau$ is normal to~$\Sigma$.
The flow lines through $s\in\Sigma$ correspond to trajectories $X(s,\tau)\in \Sigma_\varepsilon$, $0\leq \tau
\leq\varepsilon$, transverse to~$\Sigma$.

\item[$2.$]
If $\star_\Sigma$ denotes the Hodge operator defined in~$\Sigma$, and if $X_\Sigma\colon \Sigma\rightarrow\Sigma_\varepsilon$
stands for the inclusion $X_\Sigma(\cdot):=X(\cdot,0)$ then
\begin{gather*}
\star_\Sigma X^*_\Sigma(\varphi)=X^*_\Sigma(\star\varphi),
\qquad
\forall\, \varphi\in\Omega^k(\Sigma_\varepsilon).
\end{gather*}

\item[$3.$]
If $\mathcal{L}_\cdot$ denotes the Lie derivative, then
\begin{gather*}
X^*_{\Sigma}\left(\mathcal{L}_{\partial_\tau}\star\cdot\right)=
X^*_{\Sigma}\left(\star\mathcal{L}_{\partial_\tau}\cdot\right)= \star_\Sigma
X^*_{\Sigma}\left(\mathcal{L}_{\partial_\tau}\cdot\right).
\end{gather*}

\item[$4.$]
$X_\Sigma^*(\mathcal{L}_{\partial_\tau}(d^\star{\varphi}))=X_\Sigma^*(d^\star(\mathcal{L}_{\partial
\tau}{\varphi}))$, for any $\varphi\in\Omega^k(\Sigma_\varepsilon)$.

\item[$5.$]
Suppose that $\overline{\varphi}\in\Omega^1(\Sigma_\varepsilon)$ satisf\/ies $\iota_{\partial_\tau}\overline{\varphi}=0$
then
\begin{gather*}
\star_\Sigma X^\star_\Sigma(\mathcal{L}_{\partial_\tau}\overline{\varphi})=X^\star_\Sigma(\star d\overline{\varphi}).
\end{gather*}

\item[$6.$]
If $X_\Sigma^*(d^\star{\varphi})=0$, then $X_\Sigma^*(d^\star\mathcal{L}_{\partial_\tau}{\varphi})=0$, for any
$\varphi\in\Omega^k(\Sigma_\varepsilon)$.

\end{enumerate}

{\rm  B.~Corners case:} Let~$\Sigma$ be a~compact $(n-1)$-manifold with corners that is a~component of the boundary of
a~Riemannian manifold with corners~$\Sigma_\varepsilon$ diffeomorphic to a~regular cylin\-der~$\widehat{\Sigma}_\varepsilon$,~\eqref{eqn:regcyl}, provided with a~Riemannian metric~$h$.
Then there exists an embedding
\begin{gather*}
X\colon \ \widehat{\Sigma}_\varepsilon\rightarrow \Sigma_\varepsilon
\end{gather*}
such that all previous assertions hold.
\end{lma}

\begin{proof}
[Proof of case A] Consider the exponential map $Y\colon \Sigma\times[0,\varepsilon]\rightarrow \Sigma_\varepsilon$,
$Y^t(\cdot):=Y(\cdot,t)$. on a~tubular neighborhood $\Sigma_\varepsilon$ of~$\Sigma$ (see for instance~\cite{Mi}).
This means that for every initial condition $s\in\Sigma$ and $t\in[0,\varepsilon]$, $Y^t(s)\in \Sigma_\varepsilon$, is
a~geodesic passing trough $s=Y^0(s)$ whose arc-length is~$t$.
The initial velocity vector f\/ield $\frac{\partial Y^t(s)}{\partial t}|_{t=0}=\frac{\partial Y^0(s)}{\partial t}$, $s\in \Sigma$,
is a~vector f\/ield $\partial_\tau$, normal to $\Sigma\subset \Sigma_\varepsilon$.

Let $\lambda\in\Omega^{n-1}(\Sigma_\varepsilon)$ be the $(n-1)$-volume form associated to the Riemannian metric in
$\Sigma_\varepsilon$, recall that $\dim \Sigma_\varepsilon=n$.
Def\/ine $\lambda^t:=(Y^t)^*\lambda$ as the form induced by the restriction of the $(n-1)$-volume form on the embedded
$(n-1)$-hypersurface $Y^t(\Sigma)\subset \Sigma_\varepsilon$.
Now take the dif\/ferentiable function $c(t):={\int_\Sigma \lambda^0}/{\int_\Sigma\lambda^t}\in\mathbb{R}^+, $ $\forall\,
t\in [0,\varepsilon]$.
Notice that $c(0)=1$.
Then by the compactness of $\Sigma\subset\partial M$, $[c(\tau)\lambda^\tau]=[\lambda^0]\in H^{n-1}_{dR}(\Sigma)$,
for every f\/ixed $\tau\in[0,\varepsilon]$.
Hence by Moser's trick, see~\cite{Mo}, there exists an isotopy of the identity, $Z\colon \Sigma\times[0,\tau]\rightarrow
\Sigma$ such that $(Z^\tau)^*(c(\tau)\cdot\lambda^\tau)=\lambda^0$, $Z^0(s)=s$, $\forall\, s\in\Sigma$, where
$Z^t(s):=Z(s,t)$.

We def\/ine
\begin{gather*}
X(s,\tau):=Z^\tau\circ Y^\tau(s),
\qquad
\forall\,(s,\tau)\in\Sigma\times [0,\varepsilon]
\end{gather*}
also $X^t(\cdot):=X(\cdot,t)$, $X_\Sigma(\cdot):=X(\cdot,0)=X^0(\cdot)$.

Consider the explicit form of the Hodge star operator, $\star$, for the Riemannian metric~$h$ on $\Sigma_\varepsilon$,
and the star operator, $\star_\Sigma$, for the induced metric $\overline{h}:=X_\Sigma^*h$ on~$\Sigma$.
Take a~$k$-form $\varphi\in\Omega^k(\Sigma_\varepsilon)$.
If we consider a~coordinate chart $(x_1,\dots,x_{n-1})$ in~$\Sigma$.
Then locally $X^*_\Sigma(\star \varphi)$ equals the pullback of the~$k$-form
\begin{gather*}
\star\left(\sum\limits_{I} a_{I}dx^{i_1}\wedge \dots\wedge dx^{i_k}+ \sum\limits_{I'} b_{I'}dx^{i'_1}\wedge \dots\wedge
dx^{i'_{k-1}}\wedge d\tau\right)
\\
\qquad {}=\sqrt{ |\det (h_{ij} ) |}\left(\sum\limits_{J} h^{i_1j_1}\cdots h^{i_kj_k}a_{I}dx^{j_1}\wedge
\dots\wedge dx^{j_{n-k-1}}\wedge d\tau\right)
\\
\qquad \quad{}
+\sqrt{ |\det (h_{ij} ) |}\left(\sum\limits_{J'} h^{i_1'j_1'}\cdots
h^{i_k'j_{n-k}'}b_{I'}dx^{j_1'}\wedge \dots\wedge dx^{j_{n-k}'}\right).
\end{gather*}
 Here the indexes denote ordered sets $ I=\{i_1<\dots< i_{k}\}$, $J=\{j_1<\dots< j_{n-k-1}\}$ such that their union $I\cup J$, as an ordered set, corresponds to a~basis $(dx_1,\dots ,dx_{n-1})$ of $1$-forms on~$\Sigma$.
The ordered sets $I'=\{i_1'<\dots< i_{k-1}'\}$, $J=\{j_1'<\dots <j_{n-k}'\}$ are constructed in a~similar way.
Thus
\begin{gather*}
X^*_\Sigma(\star\varphi)= \sqrt{|\det(h_{ij})|}\left(\sum\limits_{J'} h^{i_1'j_1'}\cdots
h^{i_k'j_{n-k}'}b_{I'}dx^{j_1'}\wedge \dots\wedge dx^{j_{n-k}'}\right).
\end{gather*}
Meanwhile
\begin{gather*}
\star_\Sigma
X^*_\Sigma(\varphi)=\sqrt{\big|\det\big(\overline{h}_{ij}\big)\big|}\left(\sum\limits_{\{j_1'<\dots<j_{n-k}'\}}
\overline{h}^{i_1'j_1'}\cdots \overline{h}^{i_k'j_k'}b_{I'}dx^{j_1'}\wedge \dots\wedge dx^{j_k'}\right).
\end{gather*}
But $\overline{h}^{ij}=h^{ij}$ for $i,j\in\{1,\dots,n-1\}$, and also $h^{in}=\delta_{i,n}$, the Kronecker delta, since $\partial_\tau$ is
normal to~$\Sigma$.
Hence $\sqrt{|\det(\overline{h}_{ij})|}=\sqrt{|\det({h}_{ij})|}$, and
$\star_\Sigma X^*_\Sigma(\varphi)=X^*_\Sigma(\star\varphi)$.
This proves assertion~2.

If the volume form on~$\Sigma$ in local coordinates can been described as
$|\det(\overline{h}_{ij})|^{1/2}dx^1\wedge dx^2\wedge\dots\wedge dx^{n-1}$, then
$(X^\tau)^*(c(\tau)\lambda^\tau)=\lambda^0$, implies
\begin{gather*}
c(\tau) \sqrt{\big|\det\big({h}\circ X^\tau\big)_{ij}\big|} dx^1\wedge\dots \wedge dx^{n-1}=
\big|\det\big(\overline{h}_{ij}\big)\big|^{1/2}dx^1\wedge dx^2\wedge\dots\wedge dx^{n-1}.
\end{gather*}
Furthermore
$(X^\tau)^*[\mathcal{L}_{\partial_\tau}(c(\tau)\lambda)]=\frac{\partial}{\partial\tau}(X^\tau)^*(c(\tau)\lambda)$,
then
\begin{gather*}
X^*_\Sigma\left[\mathcal{L}_{\partial_\tau}(c(\tau)\lambda)\right]=
\frac{\partial}{\partial\tau}(X^\tau)^*\left(c(\tau)\lambda\right)|_{\tau=0}=\frac{\partial}{\partial\tau}\left(\lambda^0\right)|_{\tau=0}=0.
\end{gather*}
Hence
\begin{gather*}
\frac{\partial}{\partial \tau} \left(c(\tau) \sqrt{\big|\det\big({h}\circ X^\tau\big)_{ij}\big|} dx^1\wedge\dots
\wedge dx^{n-1}\right)\big|_{\tau=0}
\\
{}=\left[\frac{\partial}{\partial \tau} \big|\det\big({h}\circ X^\tau\big)_{ij}\big|^{1/2} \cdot c(t)
+\big|\det\big({h}\circ X^\tau\big)_{ij}\big|^{1/2} \frac{\partial c(\tau)}{\partial \tau}\right]\big|_{\tau=0}
dx^1\wedge\dots \wedge dx^{n-1}=0.
\end{gather*}
Recall that $c(\tau)=|\det(\overline{h}^{ij})|^{1/2}/|\det(h\circ X^\tau)|^{1/2}$; hence
\begin{gather*}
\frac{\partial c(\tau)}{\partial \tau}\Big|_{\tau=0}=
\frac{-3}{2}\big|\det\big(\overline{h}_{ij}\big)\big|^{1/2}\big|\det\big({h}\circ
X^\tau\big)_{ij}\big|^{-3/2}\frac{\partial}{\partial \tau} \big|\det\big({h}\circ
X^\tau\big)_{ij}\big|\big|_{\tau=0}.
\end{gather*}
Therefore
\begin{gather*}
\frac{\partial}{\partial \tau} \big|\det\big({h}\circ X^\tau\big)_{ij}\big|^{1/2}\big|_{\tau=0}=\frac{\partial
c(\tau)}{\partial \tau}\Big|_{\tau=0}=0.
\end{gather*}
Hence the derivative of $Z^0$ at~$\Sigma$ equals $Z^0_\star=\mathrm{Id}$, since $\frac{\partial c(\tau)}{\partial
\tau}|_{\tau=0}=0$.
Therefore{\samepage
\begin{gather*}
\frac{\partial X^\tau}{\partial\tau}|_{\tau=0}=Z^0_*\left(\frac{\partial Y^0}{\partial\tau}\right)=\partial_\tau.
\end{gather*}
This proves assertion~1.}

Now, since $\partial_\tau$ is normal to~$\Sigma$,
\begin{gather*}
\frac{\partial}{\partial\tau}\big|\det ({h} )_{ij}\big|^{1/2}\big|_{\tau=0}=
\frac{\partial}{\partial\tau}\big|\det\big(\overline{h}\big)_{ij}\big|^{1/2}\big|_{\tau=0} =0,
\end{gather*}
then $X^*_\Sigma(\mathcal{L}_{\partial_\tau}(\star \varphi))$ equals
\begin{gather*}
\sqrt{\big|\det\big(\overline{h}_{ij}\big)\big|}\left(\sum\limits_{\{j_1'<\dots<j_{n-k}'\}}
\frac{\partial}{\partial\tau} \big({h}^{i_1'j_1'}\cdots h^{i_{n-k}'j_{n-k}'}b_{I'}\big)\big|_{\tau=0}\right)
dx^{j_1'}\wedge \dots\wedge dx^{j_{n-k}'}.
\end{gather*}
Recall that the derivative of the exponential map $Y^t$ at $s\in\Sigma$, $Y^0_*\colon T_0(T_s\Sigma_\varepsilon)\simeq
T_s\Sigma_\varepsilon\rightarrow T_s\Sigma_\varepsilon$, equals the identity, $Y^0_*=\mathrm{Id}$.
This in turn implies that
\begin{gather*}
\frac{\partial}{\partial\tau}\big({h}^{i_1'j_1'}\cdots h^{i_{n-k}'j_{n-k}'}\big)\big|_{\tau=0}=0.
\end{gather*}
Therefore $X^*_\Sigma\mathcal{L}_{\partial_\tau}(\star \varphi)$ equals
\begin{gather*}
\sqrt{\big|\det\big(\overline{h}_{ij}\big)\big|}\!\left(\!\sum\limits_{\{j_1'<\dots<j_{n-k}'\}}\!\!\!\!\! {h}^{i_1'j_1'}\cdots
h^{i_{n-k}'j_{n-k}'}\frac{\partial}{\partial\tau}  (b_{I'} )|_{\tau=0}dx^{j_1'}\wedge \dots\wedge
dx^{j_{n-k}'}\!\right)\!=
X^*_\Sigma (\star\mathcal{L}_{\partial_\tau} (\varphi ) ).
\end{gather*}
This proves assertion~3.
Assertion~4 is an immediate consequence of assertion~3, and assertion~6 is in turn
a~consequence of assertion~4.

Part~5 is a~direct calculation for if $\iota_{\partial_\tau}\overline{\varphi}=0$,
$\overline{\varphi}\in\Omega^1(\Sigma_\varepsilon)$, then locally
$\overline{\varphi}=\sum\limits_{i=1}^{n-1}f_i(x,\tau)dx^i$, thus $X^*_{\Sigma}(\star d\varphi)$ equals
\begin{gather*}
X^*_\Sigma\star\left(\sum\limits_{i=1}^{n-1}\left(\sum\limits_{j\neq i}\partial_{j}f_i(x,\tau)dx^j\wedge dx^i\right) +
\sum\limits_{i=1}^{n-1}\frac{\partial}{\partial \tau}f_i(x,\tau)d\tau\wedge dx^i\right)
\\
\qquad{}= |\det h_{ij} |^{1/2}\sum\limits_{i=1}^{n-1}(-1)^{i+1} h^{1,i}\cdots\hat{h^{i,i}}\cdots
h^{n,i}\frac{\partial}{\partial \tau}f_i(x,\tau)dx^1\wedge\dots\wedge \hat{dx}^i\wedge\dots\wedge dx^{n-1},
\end{gather*}
where $\hat{h^{i,i}}$, $\hat{dx}^i$ denote missing terms.
This last expression corresponds to $\star_\Sigma X^*_\Sigma(\mathcal{L}_{\partial_\tau}\overline{\varphi})$,
therefore assertion~5 holds.
\end{proof}

\begin{proof}
[Proof of case B] Now we consider~$\Sigma$ as a~manifold with corners and $\partial\Sigma\neq\varnothing$.
We consider the exponential map $Y^{t}(s)$ for all $0\leq t\leq\epsilon(s)$, where $\epsilon^{-1}(0)=\Sigma$, recall the
def\/inition of a~regular cylinder $\widehat{\Sigma}_\varepsilon$ in~\eqref{eqn:regcyl}.

Then we use Dacorogna--Moser's argument for manifolds with boundary.
This result is proved in~\cite[Theorem~7]{DaM} for domains $\Sigma\subset\mathbb{R}^{n-1}$ with smooth boundary $\partial
\Sigma$, and for Lipschitz bounda\-ries~$\partial \Sigma$. 
In the case of general manifolds with corners~$\Sigma$, the same results holds, see Remark~2.4 and Theorem~2.3
in~\cite{ArM}.
This proves the case of regions with corners.

Since $\epsilon\colon \Sigma\rightarrow[0,1]$ in equation~\eqref{eqn:regcyl} is a~smooth increasing function, then the $(n-1)$-volume
form $c(\tau)\cdot \lambda^\tau$ coincides with the volume form $\lambda^0$ of~$\Sigma$ along $\partial \Sigma$.
They def\/ine the same cohomology form $[\lambda^0]\in H^{n-1}_{dR}(\Sigma,\partial \Sigma)$.
Now by Dacorogna--Moser, there exists $Z\colon \Sigma\times[0,\tau]\rightarrow\Sigma$, such that
$(Z^\tau)^*({\lambda}^\tau)=\lambda^0$.
Now we consider
\begin{gather*}
X^\tau=Z^\tau\circ Y^\tau(s),
\qquad
\forall\,(s,\tau)\in\Sigma\times[0,\epsilon(s)\varepsilon]
\end{gather*}
Recall that $\Sigma^\epsilon:=\epsilon^{-1}(1)\subset\Sigma$ is a~smooth deformation retract
$\Sigma^\epsilon\subset\Sigma$.
All statements for the case A~remain valid in $\Sigma^\epsilon$, since they depend on local coordinate arguments.

Take a~sequence $\varepsilon\rightarrow0$ and the corresponding regular cylinders
$\widehat{\Sigma}_\varepsilon$,~\eqref{eqn:regcyl}.
There are smooth increasing functions $\epsilon_\varepsilon\colon \Sigma\rightarrow[0,1]$, such
that for every $\varepsilon_1<\varepsilon_2$, the corresponding deformation retracts,
$\Sigma^{\epsilon_{\varepsilon_i}}=\epsilon_{\varepsilon_i}^{-1}(1)$, $i=1,2$, may be contained one in another
$\Sigma^{\epsilon_{\varepsilon_2}}\subset\Sigma^{\epsilon_{\varepsilon_1}}$.
Recall that $\epsilon_{\varepsilon_i}^{-1}(0)=\partial\Sigma$.
Therefore,~$\Sigma$ can be obtained as the closure of $\bigcup_{\varepsilon>0}\Sigma^{\epsilon_\varepsilon}$.
Hence the statements for case A~remain valid for the whole domain~$\Sigma$.
\end{proof}

\begin{dfn}
The following expression corresponds to the presymplectic structure in $\tilde{L}_\Sigma$, for the Yang--Mills action,
see for instance~\cite{Wo},
\begin{gather}
\label{eqn:presymplectic}
\tilde{\omega}_\Sigma\big(\tilde{\eta},\tilde{\xi}\big)= \frac{1}{2}\int_\Sigma X^*_\Sigma\big(\eta\wedge
d^\star\xi-\xi\wedge d^\star\eta\big),
\end{gather}
for all $\tilde{\xi},\tilde{\eta}\in\tilde{L}_{\Sigma}$ with representatives $\xi,\eta\in L_{\Sigma_\varepsilon}$.
\end{dfn}

In addition, the degeneracy subspace of the presymplectic form is
\begin{gather*}
{K}_{\omega_\Sigma}:=\big\{\tilde{\eta}\in \tilde{L}_\Sigma \,|\, {\eta}=d f, f(s,0)=0, f\in\Omega^0(\Sigma_\varepsilon),
\forall\, s\in\Sigma\big\}.
\end{gather*}
From this very def\/inition we have that the degeneracy gauge symmetries group $K_{\omega_\Sigma}$ is a~(normal) subgroup
of the identity component group $\tilde{G}_\Sigma^0\leq \tilde{G}_\Sigma$ of the gauge symmetries,
\begin{gather*}
K_{\omega_\Sigma}\trianglelefteq \tilde{G}_\Sigma^0\leq \tilde{G}_\Sigma.
\end{gather*}
Let
\begin{gather}
\label{eqn:gaugefinxing}
\Phi_{\tilde{A}_\Sigma}:=\big\{\tilde{\eta}\in \tilde{L}_\Sigma \,|\, \iota_{\partial_\tau}\eta=0, \eta\in
L_{\Sigma_\varepsilon}\ \text{reprentative of}\ \tilde{\eta}\big\}
\end{gather}
be the \emph{axial gauge fixing subspace} of $\tilde{L}_\Sigma$.
The following statement leads to a~simpler expression for the presymplectic structure.

\begin{lma}
\label{pro:1}
For every $\varphi\in L_{\Sigma_\varepsilon}$ corresponding to a~solution, there is the gauge orbit representative
\begin{gather}
\label{eqn:barphi}
\overline{\varphi}=\varphi+df,
\end{gather}
such that $\iota_{\partial\tau}\overline{\varphi}=0$, and $f|_\Sigma=0$.
\begin{enumerate}\itemsep=0pt
\item[$a)$]
Every ${K}_{\omega_\Sigma}$-orbit in $\tilde{L}_\Sigma$ intersects in just one point the subspace
$\Phi_{\tilde{A}_\Sigma}$.

\item[$b)$] The presymplectic form $\tilde{\omega}_\Sigma$ restricted to the subspace $\Phi_{\tilde{A}_\Sigma}$ may be
written as
\begin{gather*}
\tilde{\omega}_\Sigma\big(\tilde{\eta},\tilde{\xi}\big)= \frac{1}{2}\int_\Sigma
X^*_\Sigma\big(\overline{\eta}\wedge\star \mathcal{L}_{\partial_\tau}\overline{\xi}-\overline{\xi}\wedge
\star\mathcal{L}_{\partial_\tau}\overline{\eta}\big),
\end{gather*}
for every $\tilde{\xi},\tilde{\eta}\in \tilde{L}_\Sigma$ with representatives $\xi,\eta\in L_{\Sigma_\varepsilon}$.
Hence $\tilde{\omega}_\Sigma$ is a~non-degenerate $2$-form when restricted to the gauge fixing subspace
$\Phi_{\tilde{A}_\Sigma}\subset \tilde{L}_\Sigma$.
\end{enumerate}
\end{lma}

\begin{proof}[Proof of a)] Let $\big(\sum\limits_{i=1}^{n-1}\eta^idx^i\big)+\eta^\tau d\tau$ be a~local expression for a~solution
$\eta\in L_{\Sigma_\varepsilon}$.
Let us apply a~gauge symmetry
\begin{gather*}
X^*_\Sigma\left(\eta+df\right) =\sum\limits_{i=1}^{n-1}\big(\eta^i+\partial_if\big)dx^i+\big(\eta^\tau+\partial_\tau f\big)d\tau
\end{gather*}
in such a~way that $\eta^\tau+\partial_\tau f=0$.
We can solve the corresponding ODE for $f(s,\tau)$ once we f\/ix an initial condition $f(s,0)=g(s)$.
If we take this initial condition~$g(s)$ as a~constant, then we get a~gauge symmetry in $K_{\omega_\Sigma}$.
The remaining part is a~straightforward calculation.
This proves~a).
The other assertion may be inferred from Lemma~\ref{pro:1}.
\end{proof}

This result shows that Axioms~\ref{ax:2} and~\ref{ax:3} are satisf\/ied.
Let
\begin{gather*}
A_\Sigma:=\tilde{A}_\Sigma/K_{\omega_\Sigma}, L_{\Sigma}=\tilde{L}_\Sigma/K_{\omega_\Sigma}
\end{gather*}
be the quotients by the linear space $K_{\omega_\Sigma}$ corresponding to degenerate gauge symmetries.
And also let $G_\Sigma^0:=\tilde{G}^0_\Sigma/K_{\omega_\Sigma}$ be the quotient by the normal subgroup.
By Lemma~\ref{pro:1}, when we restrict the quotient class $\xymatrix{\tilde{A}_\Sigma\ar@{->>}[r] &A_\Sigma}$ to
$\Phi_{\tilde{A}_\Sigma}$, then we get an isomorphism of af\/f\/ine spaces.
Let $\omega_\Sigma$ be the corresponding \emph{symplectic structure} on $A_\Sigma$ induced by the restriction of
$\tilde{\omega}_\Sigma$ to the subspace $\Phi_{\tilde{A}_\Sigma}\subset \tilde{L}_\Sigma$.

We now proceed to give a~precise description of the symplectic space $L_\Sigma$.

Lemma~\ref{lma:embeddingndim} implies that
\begin{gather}
\label{eqn:Xwedge}
X_\Sigma^*\big(\overline{\xi}\wedge\star\mathcal{L}_{\partial_\tau}\overline{\eta}		\big)=
X_\Sigma^*\big(\overline{\xi}\big)\wedge\star_\Sigma X^*_\Sigma\big(\mathcal{L}_{\partial_\tau}\overline{\eta}\big),
\end{gather}
where $\star_\Sigma$ stands for the Hodge star on~$\Sigma$.
Since
$\iota_{\partial_\tau}(\mathcal{L}_{\partial_\tau}\overline{\eta})=\iota_{\partial_\tau}(\iota_{\partial_\tau}d\overline{\eta})=0$,
then we have a~linear map $L_{\Sigma_\varepsilon}\rightarrow \Omega^1(\Sigma)\times\Omega^1(\Sigma)$, where
\begin{gather}
\label{eqn:eta->phi}
{\eta}\mapsto
\big(\phi^\eta,\dot{\phi^\eta}\big):=\big(X_\Sigma^*(\overline{\eta}),X_{\Sigma}^*\big(\mathcal{L}_{\partial_\tau}\overline{\eta}\big)\big),
\end{gather}
for every $\tilde{\eta}\in \tilde{L}_\Sigma$ with representative $\xi,\eta\in L_{\Sigma_\varepsilon}$ and
$\overline{\eta}$ def\/ined in~\eqref{eqn:barphi}.
that leads to a~map
\begin{gather}
\label{eqn:A_Sigma}
L_{\Sigma}\rightarrow\Omega^1(\Sigma)\times\Omega^1(\Sigma)\simeq 	T\big(\Omega^1(\Sigma)\big),
\end{gather}
where we consider the identif\/ication with the tangent space $T(\Omega^1(\Sigma))$.

Notice that $\iota_{\partial_\tau}\overline{\eta} =0$ implies that $\eta\in L_{\Sigma_\varepsilon} $ corresponds to
a~$1$-form $\phi^{\eta}$ on~$\Sigma$.
Notice also that $d^\star d\eta=0$ implies
\begin{gather*}
d^\star (\iota_{\partial_\tau}d\overline{\eta} )=d^\star (\mathcal{L}_{\partial_\tau}\overline{\eta} )=0.
\end{gather*}
Hence $d^{\star_\Sigma}(X_\Sigma^*(\mathcal{L}_{\partial_\tau}\overline{\eta}))=0$.
Therefore $\dot{\phi}^\eta\in\ker d^{\star_\Sigma}$.

We have the following expression for the symplectic structure on $L_\Sigma$
\begin{gather}
\label{eqn:symplectic}
{\omega}_\Sigma\big(\big(\phi^{\eta},\dot{\phi}^\eta\big),\big(\phi^{\xi},\dot{\phi}^{\xi}\big)\big)= \frac{1}{2}\int_\Sigma
\big(\phi^\eta\wedge\star_\Sigma \dot{\phi^\xi}-\phi^{\xi}\wedge \star_\Sigma\dot{\phi^{\eta}}\big),
\end{gather}
for every $(\phi^\xi,\dot{\phi}^\xi)$, $(\phi^\eta,\dot{\phi}^\eta)\in L_\Sigma$,
with representatives $\overline{\xi}$, $\overline{\eta}\in L_{\Sigma_\varepsilon}$.
From this very def\/inition we can verify Axiom~\ref{ax:4}, i.e., translation invariance and also relation~\eqref{eqn:2} where
\begin{gather*}
\big[\big(\phi^{\eta},\dot{\phi}^\eta\big),\big(\phi^{\xi},\dot{\phi}^{\xi}\big)\big]_\Sigma:=\int_\Sigma\phi^\eta\wedge\star_\Sigma\dot{\phi^\xi}.
\end{gather*}
Furthermore Axiom~\ref{ax:6} is easily verif\/ied and the claims from Axiom~\ref{ax:5} can be inferred from the relation
$\star_{\overline{\Sigma}}=-\star_\Sigma$.

With this result we f\/inish the kinematical part of the axiomatic description, i.e., Axioms~\ref{ax:1}--\ref{ax:6}.

\subsection{Symplectic reduction}

In this subsection we assume also two cases as in the previous subsection,
either:
\begin{enumerate}\itemsep=0pt
\item[A.] {\it $\Sigma$ is a~smooth closed $(n-1)$-dimensional manifold}, or
\item[B.] {\it $\Sigma$ is an $(n-1)$-dimensional manifold with corners.}
\end{enumerate}

We still need to describe the quotient for the symplectic action of the gauge group~$G^0_\Sigma$ on~$L_\Sigma$.
The suitable gauge f\/ixing space $\Phi_\Sigma$ in $L_\Sigma$ for this action will be the space of divergence free
$1$-forms, i.e., we def\/ine
\begin{gather*}
\Phi_{A_\Sigma}:=\big\{\big(\phi,\dot{\phi}\big)\in L_\Sigma \,|\,
d^{\star_\Sigma}\phi=0=d^{\star_\Sigma}\dot{\phi}\big\}.
\end{gather*}

The following task is the detailed description of the symplectic quotient space
\begin{gather*}
L_\Sigma/G_\Sigma^0\simeq\Phi_{A_\Sigma}.
\end{gather*}

{\it A.~Smooth case.} We recall some useful facts of Hodge--Morrey--Friedrich theory for manifolds with boundary, see for
instance~\cite{AM,Du,DuS, Sc}.
We can consider both Neumann and Dirichlet boundary conditions in order to def\/ine~$k$-forms on a~manifold~$V$, i.e.,
\begin{gather*}
\Omega^k_{N}(V):=\big\{\varphi\in\Omega^k(V)\,|\, X_{\partial V}^*(\star\varphi)=0\big\},
\qquad
\Omega^k_D(V):=\big\{\varphi\in\Omega^k(V) \,|\, X_{\partial V}^*(\varphi)=0\big\}.
\end{gather*}
The dif\/ferential~$d$ preserves the Dirichlet complex $\Omega^k_{D}(V)$ and on the other hand, the codif\/fe\-ren\-tial~$d^\star$ preserves the Neumann complex $\Omega^k_{N}(V)$.
In addition, the space $\mathfrak{H}^k(V)$ of harmonic f\/ields $d\varphi=0=d^\star\varphi$, turns out to be inf\/inite-dimensional.
Nevertheless f\/inite-dimensional spaces arise when we restrict to Dirichlet or Neumann boundary conditions
$\mathfrak{H}^k_N(V)$, $\mathfrak{H}^k_D(V)$.

According to Hodge theory~\cite{Sc}, associated with the inner product
\begin{gather*}
\int_\Sigma\phi\wedge\star_\Sigma\phi',
\qquad
\phi,\phi'\in\Omega^1(\Sigma)
\end{gather*}
we have an orthogonal decomposition $\phi=\phi_{\mathfrak{h}}+d^{\star_\Sigma}\alpha_\phi$,
\begin{gather*}
\big\{\phi\in\Omega^1(\Sigma) \,|\, d^\star\phi=0\big\}=\mathfrak{H}^1(\Sigma)\oplus d^{\star_\Sigma}\Omega^2(\Sigma),
\end{gather*}
where the space $\mathfrak{H}^1(\Sigma)$ of harmonic $1$-forms has rank $b=\dim\mathfrak{H}^1(\Sigma)$.

According to~\cite{Du,DuS}, the space of harmonic forms on a~smooth manifold~$\Sigma$ with \emph{smooth}
boundary $\partial \Sigma$, has rank
\begin{gather*}
b=\dim\mathfrak{H}_N^1(\Sigma)=\dim H_1(\Sigma)=\dim H_{n-2}(\Sigma,\partial \Sigma).
\end{gather*}

{\it B.~Corners case.} For manifolds with corners~$\Sigma$, the space of harmonic forms has the same description.
Take a~homeomorphism $F\colon \Sigma'\rightarrow\Sigma$, that def\/ines a~dif\/feomorphism, with lack of dif\/ferentiability on~$\partial \Sigma'$.
Here~$\Sigma'$ is a~smooth manifold with smooth boundary homeomorphic to~$\Sigma$.
If $\phi\in\mathfrak{H}_N^1(\Sigma)$ is a~harmonic form with null normal component, then
$F^*(\phi)|_{\partial\Sigma'}\in \mathfrak{H}_N^1(\Sigma')$ is also a~well def\/ined harmonic form on $\partial
\Sigma'$.
Hence for manifolds with corners~$\Sigma$, harmonic forms have also rank given by the Betti number.

The following lemma fulf\/ills Axiom~\ref{ax:8} and provides the gauge-f\/ixing space def\/inition required in Axiom~\ref{ax:9}.

\begin{lma}
\label{lma:intersection}
Let~$\Sigma$ be a~closed smooth manifold or a~manifold with corners of dimension $n-1$.
For $\eta\in L_{\Sigma_\varepsilon}$, take $(\phi^\eta,\dot{\phi}^\eta)\in T\Omega^1(\Sigma)$ as defined
in~\eqref{eqn:eta->phi}, with gauge transformation group~$G_\Sigma^0$.
\begin{enumerate}\itemsep=0pt

\item[$a)$] The gauge group action of $G^0_\Sigma$ on $L_\Sigma$ is induced in the tangent space $T\Omega^1(\Sigma)$~by
the translation action ${\phi}\mapsto\phi+df, f\in\Omega^0(\Sigma)$ on $\Omega^1(\Sigma)$.

\item[$b)$]
Every ${G}^0_{\Sigma}$-orbit in ${L}_\Sigma$ intersects the subspace $\Phi_{{A}_\Sigma}$ in just one point .

\item[$c)$] The symplectic form ${\omega}_\Sigma$ is preserved under the $G^0_\Sigma$-action.
\end{enumerate}
\end{lma}

\begin{proof}[Proof of b)]
Consider $X^*_\Sigma\eta=\sum\limits_{i=1}^{n-1}\eta^idx^i$, a~local expression for a~solution $\eta\in
L_{\Sigma_\varepsilon}\cap \Phi_{\tilde{A}_{\Sigma}}\subset L_{\Sigma_\varepsilon}$.
Consider $\overline{f}\colon {\Sigma}\rightarrow \mathbb{R}$, then
$d^{\star_\Sigma}(X^*_\Sigma(\eta)+d\overline{f})=0$ implies
\begin{gather}
\label{eqn:ODE12}
\sum\limits_{i=1}^{n-1}\partial_ i\left[\big|\det\big(\overline{h}\big)\big|^{1/2} \sum\limits_{i=1}^{n-1}\big(\eta^i+\partial_i
\overline{f}\big)(-1)^ih^{1,i}  \cdots \hat{h}^{i,i} \cdots  h^{i,n-1}\right]=0.
\end{gather}
The existence and regularity of a~solution, $\overline{f}(s)$, for this PDE on~$\Sigma$ is warranted precisely by Hodge
theory.
Since
\begin{gather*}
X^*_\Sigma(\eta)\in\Omega^1(\Sigma)\simeq d\Omega^0(\Sigma)\oplus		\mathfrak{H}^1(\Sigma)\oplus
d^{\star_\Sigma}\Omega^2(\Sigma),
\end{gather*}
there exists $\overline{f}\in\Omega^0(\Sigma)$, such that $X^*_\Sigma(\eta) + d\overline{f}$ is the orthogonal
projection of $	X^*_\Sigma(\eta)$ onto $\ker d^{\star_\Sigma}\simeq\mathfrak{H}^1(\Sigma)\oplus
d^{\star_\Sigma}\Omega^2(\Sigma)$.
Def\/ine
\begin{gather*}
\phi^\eta:=X^*_\Sigma(\eta)+d\overline{f}\in\ker d^{\star_\Sigma}.
\end{gather*}
On the other hand
\begin{gather}
\label{eqn:LL}
d^{\star}\mathcal{L}_{\partial_\tau} (\eta+df )=0
\end{gather}
implies
\begin{gather}
\label{eqn:ODE22}
\sum\limits_{i=1}^{n-1}\partial_ i\left[|\det({h})|^{1/2}
\sum\limits_{i=1}^{n-1}\big(\partial_\tau\eta^i+\partial_\tau\partial_i {f}\big) (-1)^ih^{1,i} \cdots
\hat{h}^{i,i} \cdots  h^{i,n-1}\right]=0.
\end{gather}
When we substitute $\partial_\tau\eta^i+\partial_\tau\partial_i {f}$ by the coef\/f\/icients $\phi^i_\tau$, of a~time
dependent $1$-form in~$\Sigma$, $\phi_\tau\in\Omega^1(\Sigma)$, equation~\eqref{eqn:ODE22} has a~solution $\phi_\tau$.
This leads to an ODE for $g_i(s,\tau):=\partial_if$,
\begin{gather}
\label{eqn:ODE3}
\partial_\tau\eta^i+\partial_\tau\partial_i {f}=\phi^i_\tau.
\end{gather}
Equation~\eqref{eqn:ODE3} can be solved once we f\/ix the boundary condition
$\partial_if(x^i,0)=\partial_i\overline{f}(x^i)$.
This boundary condition, in turn, has been obtained by solving~\eqref{eqn:ODE12} in~$\Sigma$.

We conclude that $\sum\limits_{i=1}^{n-1}g_i(s,\tau)dx^i$ is an exact form on~$\Sigma$, so that there exists
$f(s,\tau)\in\Omega^0(\Sigma_\varepsilon)$ such that~\eqref{eqn:LL} holds.

To conclude def\/ine $\dot{\phi}^\eta:=X^*_\Sigma(\mathcal{L}_{\partial_\tau}(\eta+df))$, notice
that $(\phi^\eta,\dot{\phi}^\eta)\in\Phi_{A_\Sigma}$.

Remark that from the very form of the solution $\phi_\tau=X^*_\Sigma(\eta_\tau)+d\overline{f}_\tau$, $\phi_\tau$ and
$\partial_\tau\eta$ have the same integrals along closed cycles, hence they have the same cohomology class in
$\mathfrak{H}^1(\Sigma)$.
\end{proof}

The axial Gauge f\/ixing space $\Phi_\Sigma$ can be described with
\begin{gather*}
 T\big[\mathfrak{H}^1(\Sigma)\oplus d^{\star_\Sigma}\Omega^2(\Sigma)\big]\simeq
\big[T\mathfrak{H}^1(\Sigma)\big]\times \big[T(d^{\star_\Sigma}\Omega^2(\Sigma))\big],
\end{gather*}
where
we take tangent spaces.
Recall that according to Hodge theory the space of coclosed $1$-forms can be described
as $\mathfrak{H}^1(\Sigma)\oplus d^{\star_\Sigma}\Omega^2(\Sigma)$.
In the abelian case the holonomy $\mathrm{hol}_\gamma(\phi)=\exp\oint_\gamma\phi\in G$ of a~connection~$\phi$ along
a~closed trajectory~$\gamma$ can be def\/ined up to cohomology class of~$\gamma$.
Recall that for $G=U(1)$, $\int_\gamma\phi\in\sqrt{-1}\mathbb{R}$.
Thus by considering independent generators $\{\gamma_1,\dots, \gamma_{b}\}$ of the homology $H_1(\Sigma)$, and a~dual
harmonic basis $\phi_\mathfrak{h}^1,\dots,\phi_\mathfrak{h}^{b}$ we have the exact sequence
\begin{gather*}
\xymatrix{
0\ar[r]&\oplus_{i=1}^{b}\mathbb{Z}\cdot\big[\phi^i_{\mathfrak{h}}\big]\ar[r]&\mathfrak{H}^1(\Sigma)\ar[r]^{\mathrm{hol}_{\gamma_{i}}}&G^{b}\ar[r]&1.
}
\end{gather*}
Hence there is a~surjective map by the dif\/ferential $\mathrm{Dhol}_\Gamma\colon T\mathfrak{H}^1(\Sigma)\rightarrow TG^{b}$.

Now we consider the reduction of $\Phi_{A_\Sigma}$ under the action of the discrete group $G_\Sigma/G^0_\Sigma$.

\begin{lma}
\label{tma:1}
Let~$\Sigma$ be a~$(n-1)$-dimensional smooth manifold or a~manifold with corners.
We have the quotient space
\begin{gather*}
A_\Sigma/G_\Sigma = \Phi_{A_\Sigma}/\big(G_\Sigma/G^0_\Sigma\big)\simeq T\big(G^{b}\big)\times
T\big(d^{\star_\Sigma}\Omega^2(\Sigma)\big)
\end{gather*}
with reduced symplectic structure $\overline{\omega_\Sigma}$ given in~\eqref{eqn:symplectic}.
\end{lma}

\subsection{Factorization on hypersurfaces}

Now we complete the symplectic reduction picture for hypersurfaces described in Axioms~\ref{ax:7}--\ref{ax:10}.
First we consider the factorization given in Axiom~\ref{ax:7}.
In this section we have denoted alternatively smooth closed manifolds or manifolds with corners as~$\Sigma$.
For the results stated in this particular subsection, the convention will be dif\/ferent:

In this subsection~$\Sigma$ will denote a~\emph{hypersurface}, i.e., a~topological manifold with a~stratif\/ied space
structure~$|\Sigma|$.

The~$r$-forms in the~$k$-skeleton, $r\leq k$, $k=0,1,2,\dots, n-1$ is the set of restrictions of these~$r$-forms to its
faces:
\begin{gather*}
\Omega^r\big(|\Sigma|^{(n-1)} \big)= \cup_{i=1}^m\big\{\varphi^i\in\Omega^k\big(\Sigma^i\big) \,|\,
\varphi^i|_{\Sigma^j}=\varphi^j \; \forall\, (i,j)\in\mathcal{P} \big\},
\\
\Omega^r\big(|\Sigma|^{(n-2)}\big)= \cup\big\{\varphi^I\Omega^{k}\Sigma^I \,|\, \varphi^I|_{\Sigma^J}=\varphi^J,\,
{I,J\in \mathcal{P}} \big\}.
\end{gather*}
Notice that we have the inclusion of~$k$-forms on stratif\/ied spaces, for $r=0,1$, given by the pullbacks
\begin{gather*}
\Omega^r(\Sigma)=\Omega^r\big(|\Sigma|^{(n-1)}\big)\subset \Omega^r\big(|\Sigma|^{(n-2)}\big).
\end{gather*}
To give a~detailed description of the space of divergence-free f\/ields on~$\Sigma$, let us f\/irst consider harmonic
f\/ields.

If $\phi_\mathfrak{h}\in \mathfrak{H}^1(\Sigma^i)$, then the restriction, $\phi^I_{\mathfrak{h}}$, over every face
closure $\Sigma^I\subset \Sigma^i$, contained in $\Sigma^i$, is harmonic as stated in the following lemma, which
consists of two parts one for regular cylinders and another one for $(n-1)$-dimensional stratif\/ied spaces.

\begin{lma}[Theorems 7 and 8 in~\cite{AGO}]
Let $\Sigma_\varepsilon$ be a~Riemannian manifold with corners homeomorphic to the regular cylinder
$\widehat{\Sigma}_\varepsilon$,~\eqref{eqn:regcyl}.
For every harmonic form $\varphi\in\mathfrak{H}^r(\Sigma_\varepsilon)$, the following are true:
\begin{enumerate}\itemsep=0pt

\item[$1.$] $\varphi$ is closed and coclosed, that is $d\varphi=0=d^{\star}\varphi$,
i.e., $\varphi\in\mathfrak{H}^r(\Sigma_\varepsilon)$.

\item[$2.$] $\varphi^i:=\varphi|_{\Sigma^i}\in\mathfrak{H}^r(\Sigma^i)$, where
${\Sigma}^i\subset|\Sigma|^{(n-1)}$ are the $(n-1)$-dimensional faces.

\item[$3.$] The boundary and coboundary operators satisfy,
\begin{gather*}
\big(\big(d\varphi\big)^i\big)= \big(d(\varphi^i)\big),
\qquad
\big(\big(d^\star\varphi\big)^i\big)=\big(d^{\star_{\Sigma^{i}}}\big(\varphi^i\big)\big),
\end{gather*}
so that they define complexes $(\Omega(|\Sigma|^{(n-2)}),d)$, $(\Omega(|\Sigma|^{(n-2)})),d^\star)$.

\item[$4.$] $\varphi|_{\Sigma^i}\in\mathfrak{H}^r_N(\Sigma^i)$, if and only if $\iota_{\partial_\tau}\varphi=0$, where
$\partial_\tau$ is a~vector field normal to $\Sigma^i$.
\end{enumerate}

Let $|\Sigma|^{(n-1)}=|\Sigma|$ be an $(n-1)$-dimensional stratified space homeomorphic to an $(n-1)$-dimensional
manifold.
For every harmonic~$r$-form $\phi\in\mathfrak{H}^r(|\Sigma|)$

\begin{enumerate}\itemsep=0pt

\item[$1.$] $\varphi$ is closed and coclosed, that is $d\phi=0=d^{\star_{\Sigma^i}}\phi$,
i.e., $\phi^i:=\phi|_{\Sigma^i}\in\mathfrak{H}^r(\Sigma^i)$, for each $(n-1)$-dimensional closed stratum $\Sigma^i$.

\item[$2.$] $\phi^I:=\phi|_{\Sigma^I}\in\mathfrak{H}^r(\Sigma^I)$, where ${\Sigma}^I\subset|\Sigma|^{(n-2)}$ are
the $(n-2)$-dimensional faces.

\item[$3.$] The boundary and coboundary operators satisfy,
\begin{gather*}
\big(\big(d\phi\big)^I\big)= \big(d\big(\phi^I\big)\big),
\qquad
\big(\big(d^\star\phi\big)^I\big)=\big(d^{\star_{\Sigma^{I}}}\big(\phi^I\big)\big),
\end{gather*}
so that they define complexes $(\Omega(|\Sigma|^{(n-2)}),d)$, $(\Omega(|\Sigma|^{(n-2)})),d^\star)$.
\end{enumerate}
\end{lma}

This f\/inishes the description of harmonic forms on the stratif\/ied space $|\Sigma|$.
Notice that when~$\Sigma^i$ are balls then the harmonic forms $\phi\in\Omega^r(\Sigma^i)$ are completely def\/ined~by
their Dirichlet boundary conditions on $\partial \Sigma^i$.

We also have an analogous of Hodge--Morrey--Friedrichs decomposition for stratif\/ied spaces, that follows also from Theorems~7 and~8
in~\cite{AGO}:

\begin{cor}\label{cor:orthogonaldecompostionwithcorners}\qquad
\begin{enumerate}\itemsep=0pt
\item[$1.$] There is an orthogonal decomposition
\begin{gather*}
\Omega^r(|\Sigma|)=\mathfrak{H}_N^r (|\Sigma| )\oplus \big(\mathfrak{H}^r(|\Sigma|)\cap
d\Omega^{r-1}(|\Sigma|)\big)\oplus
 d\Omega^{r-1}_D (|\Sigma| )\oplus d^\star\Omega_N^{r+1} (|\Sigma| ).
\end{gather*}
\item[$2.$] In particular there is an orthogonal decomposition for divergence-free fields
\begin{gather*}
\ker  \big[d^\star\colon \Omega^r (|\Sigma| )\rightarrow\Omega^{r-1} (|\Sigma| )\big]=
\mathfrak{H}^r_N (|\Sigma| )\oplus d^\star\Omega^{r+1}_N (|\Sigma|).
\end{gather*}
\end{enumerate}
\end{cor}

Therefore the divergence free $1$-forms on $|\Sigma|$ are described as
\begin{gather*}
\mathfrak{H}^1 (|\Sigma| )\oplus d^{\star_{\Sigma}}\Omega^2 (|\Sigma| ),
\end{gather*}
thus we could def\/ine the gauge-f\/ixing space as
\begin{gather*}
\Phi_{A_{\Sigma}}:=T\big(\mathfrak{H}^1 (|\Sigma| )\big)\times T\big(\star_\Sigma
d\Omega^0 (|\Sigma| )\big).
\end{gather*}
From the projections $\check{\Sigma}^i\rightarrow\Sigma^i\subset \Sigma$ we obtain the linear maps
\begin{gather*}
\mathfrak{H}^1 (|\Sigma| )\subset \oplus_{i=1}^r\mathfrak{H}^1\big(\check{\Sigma}^i \big)
\end{gather*}
and
\begin{gather*}
\Omega^0 (|\Sigma| )\subset \oplus_{i=1}^m\Omega^0\big(\check{\Sigma}^i\big).
\end{gather*}
If $\Phi_{A_{\check{\Sigma}^i}}:=T(\mathfrak{H}^1({\check{\Sigma}}^i))\oplus
T(\star_{\Sigma^i}d\Omega^0({\check{\Sigma}}^i))$, then we get the injective maps
\begin{gather*}
\Phi_{A_{\Sigma}}\rightarrow \prod_{i=1}^m \Phi_{A_{\check{\Sigma}^i}}=:\Phi_{A_{|\Sigma|^{2}}}.
\end{gather*}
This inclusion is the restriction of an inclusion referred to in Axiom~\ref{ax:7}, as is described in the following
commuting diagram
\begin{gather*}
\xymatrix{
T\big(\Omega^1 (|\Sigma| )\big)
\ar[r]
&
\prod\limits_{i=1}^m T\big(\Omega^1\big(\check{\Sigma}^i\big)\big)=T\big(\oplus_{i=1}^m \Omega^1\big(\check{\Sigma}^i\big)\big)
\\
L_{{\Sigma}}
\ar[r]\ar@{^{(}->}[u]
&
\oplus_{i=1}^m L_{{\check{\Sigma}^i}}=L_{{|\Sigma|^2}}
\ar@{^{(}->}[u]
\\
\Phi_{A_{\Sigma}}
\ar@{^{(}->}[u]
\ar[r]
&
\prod\limits_{i=1}^m \Phi_{A_{\check{\Sigma}^i}}=:\Phi_{A_{|\Sigma|^2}}
\ar@{^{(}->}[u]
}
\end{gather*}

Similarly, the inclusions in the af\/f\/ine spaces $A_{\Sigma}\rightarrow \prod\limits_{i=1}^m A_{\check{\Sigma}^i}$ can be
described.
Also for the gauge symmetries we have exact sequences
\begin{gather*}
\xymatrix{
H^0_{dR} (|\Sigma| )
\ar@{-->}[r]\ar[d]&
\mathbb{R}^m\ar[d]
\\
\Omega^0 (|\Sigma| )
\ar[d]
\ar@{-->}[r]
&
\oplus_{i=1}^m\Omega^0\big(\check{\Sigma}^i\big)
\ar[d]
\\
G_{\Sigma}^0\ar@{-->}[r]
&
\oplus_{i=1}^m G_{\check{\Sigma}^i}=:G_{|\Sigma|^{2}}
}
\end{gather*}
where $G_{\check{\Sigma}^i}\simeq\Omega^0(\check{\Sigma}^i)/\mathbb{R}$ (since~$\check{\Sigma}^i$ is simply connected),
and $G_{\Sigma}^0$ stands for the identity component of the gauge symmetries group~$G_{\Sigma}$.

We use Lemma~\ref{tma:1} to establish the following claim.

\begin{tma}
Let~$\Sigma$ be a~hypersurface with a~stratified space structure~$|\Sigma|$.
We have the gauge fixing space
\begin{gather*}
\Phi_{A_\Sigma} = \Phi_{A_\Sigma}/G^0_\Sigma\simeq T\big(\mathfrak{H}^1 (|\Sigma| )\big)\times
T\big(\star_\Sigma d\Omega^0 (|\Sigma| )\big)\subset \prod_{i=1}^m \Phi_{A_{\check{\Sigma}^i}}
\end{gather*}
with symplectic structure $\overline{\omega_\Sigma}$ induced by the pullback of
$\omega_{\check{\Sigma}^1}\oplus\dots\oplus\omega_{\check{\Sigma}^m}$.
We also have the quotient space
\begin{gather*}
A_\Sigma/G_\Sigma =  T\big(G^{b}\big)\times T\big(d^{\star_\Sigma}\Omega^2 (|\Sigma| )\big).
\end{gather*}
Where $b=\dim \mathfrak{H}^1(|\Sigma|)$ is the Betti number of~$\Sigma$,
\begin{gather*}
b=\dim H_{n-2}(\Sigma;\partial\Sigma)=\dim H_1(\Sigma),\qquad  n-1=\dim{\Sigma}.
\end{gather*}
\end{tma}

This proves the validity of the factorization Axioms~\ref{ax:7},~\ref{ax:10}.

For the smooth case let us consider a~hypersurface~$\Sigma$ as a~disjoint union of oriented hypersurfaces
${\Sigma}=\Sigma^1\sqcup\dots\sqcup \Sigma^m$.
Then there is a~linear map
\begin{gather*}
\Omega^1({\Sigma})\rightarrow \Omega^1(\Sigma^1)\oplus\dots\oplus\Omega^1(\Sigma^m)
\end{gather*}
given by $\eta\mapsto X_{\Sigma^1}^*(\eta)\oplus\dots\oplus X_{\Sigma^m}^*(\eta)$.
This map induces the isomorphism $r_{\Sigma;|\Sigma|^{n-1}}\colon L_{\Sigma}\rightarrow
L_{|\Sigma|^{n-1}}=L_{\Sigma^1}\oplus\dots\oplus L_{\Sigma^m}$.
Furthermore, the chain decomposition
\begin{gather*}
\int_{\Sigma}\cdot =\int_{\Sigma^1}\cdot+\dots+\int_{\Sigma^m}\cdot
\end{gather*}
verif\/ies Axiom~\ref{ax:7}$'$.
The proof of Axiom~\ref{ax:10}$'$ is similar.

Let us now look at the content  of Axiom~\ref{ax:8}.
Let $G_M^0 =\{df \,|\, f\in\Omega^0(M)\}$ be the identity component of the bulk gauge symmetry group $G_M$.
Each bulk symmetry $df\in G^0_M$ induces a~symmetry $X_\Sigma^*(df)\in G^0_{\partial M_\varepsilon}$ in the boundary
cylinder ${\partial M_\varepsilon}:=\cup_i \Sigma^i_\varepsilon$, and also a~symmetry $h_M(df)\in G_{\partial M}^0$ in
the boundary conditions.
This was mentioned in the locality arguments in Subsection~\ref{subsec:locality}.

There is an extension of local gauge actions: In the particular case of trivial principal bundle local gauge symmetries
in $G_{\partial M}$ extend via partitions of unity to symmetries in the bulk~$G_M$.
This means that we can def\/ine sections $\sigma: G_{\partial M}^0\rightarrow G_M$ of the homomorphism $G_M\rightarrow
G_{\partial M}$.
Hence there is a~well def\/ined (set-theoretic) orbit map
\begin{gather*}
\overline{r}_M\colon \ L_M/G_M^0\rightarrow L_{\partial M}/G_{\partial M}^0.
\end{gather*}
Furthermore, by linearity of the actions $L_M/G_M^0$, $A_M/G_M^0$ have linear and af\/f\/ine structures respectively.
This proves Axiom~\ref{ax:8}.

The kinematic description of gauge f\/ields is now completed from Axioms~\ref{ax:1}--\ref{ax:10}.

\section{Dynamics modulo gauge}
\label{sec:Dynamics}

These paragraphs are aimed to verify Axioms~\ref{ax:9}--\ref{ax:12} where dynamics of gauge f\/ields is constructed.
We discuss the behavior of the solutions near the boundary in more detail.
Recall that here is a~map $\tilde{r}_M\colon L_M\rightarrow \tilde{L}_{\partial M}$ coming from the restriction of the
solutions to germs on the boundary. Composing with the quotient class map by the space~$K_{\omega_{\partial M}}$, we have a~map $r_M\colon L_M\rightarrow
L_{\partial M}$.
Let $L_{\tilde{M}}\subset L_\Sigma$ be the image under this map.
The aim is to describe the image $L_{\tilde{M}}\subset L_{\partial M}$ of the space of solutions as a~Lagrangian
subspace once we take the gauge quotient.

We recall some useful facts of Hodge--Morrey--Friedrich theory for manifolds with boundary for~$k$-forms on~$M$.

\begin{lma}[A.~Smooth case~\cite{Sc}]
\label{lma:orthogonaldecompostion}
Suppose that~$M$ is a~smooth Riemannian manifold with boundary
\begin{enumerate}\itemsep=0pt
\item[$1.$] There is an orthogonal decomposition
\begin{gather*}
\Omega^k(M)=\mathfrak{H}_N^k(M)\oplus \big(\mathfrak{H}^k(M)\cap d\Omega^{k-1}(M)\big)\oplus d\Omega^{k-1}_D(M)\oplus
d^\star_N \Omega_N^{k+1}(M).
\end{gather*}
\item[$2.$] In particular there is an orthogonal decomposition for divergence-free fields
\begin{gather*}
\ker \big[d^\star\colon \Omega^k_N(M)\rightarrow\Omega^{k-1}_N(M)\big]=\mathfrak{H}^k_N(M)\oplus d^\star\Omega^{k+1}_N(M).
\end{gather*}

\item[$3.$] Each de Rham cohomology class can be represented by a~unique harmonic field without normal component, i.e., there
is an isomorphism
\begin{gather*}
H^k_{dR}(M)\simeq \mathfrak{H}^k_N(M).
\end{gather*}

\item[$4.$] Each de Rham relative cohomology class can be represented by a~harmonic field null at the boundary, i.e., there is
an isomorphism
\begin{gather*}
H^k_{dR}(M,\partial M)\simeq \mathfrak{H}^k_D(M).
\end{gather*}
\end{enumerate}
\end{lma}

For the corners case we consider the stratif\/ied space structure $|M|$ of the manifold with corners~$M$ in order to give
a~precise def\/inition of the spaces of Neumann and Dirichlet boundary conditions on~$k$-forms
\begin{gather*}
\Omega^k_D(|M|):=\big\{\varphi\in\Omega^k(|M|) \,|\, \varphi|_{|\partial M|}=0\big\},
\qquad
\Omega^k_N(|M|):=\big\{\varphi\in\Omega^k(|M|) \,|\, \star\varphi|_{|\partial M|}=0\big\}.
\end{gather*}
The decomposition given in Corollary~\ref{cor:orthogonaldecompostionwithcorners}, has a~corners counterpart given by the
following result.

\begin{lma}[B.~Corners case]
Suppose that~$M$ is a~Riemannian manifold with corners with stratified space structure~$|M|$.
\begin{enumerate}\itemsep=0pt
\item[$1.$] There is an orthogonal decomposition
\begin{gather*}
\Omega^k(|M|)=d\Omega^{k-1}_D(|M|)\oplus\mathfrak{H}^k_N(|M|)\oplus\big(\mathfrak{H}^k(|M|)\cap d\Omega^{k-1}(|M|)\big)\oplus
d^\star\Omega_N^{k+1}(|M|).
\end{gather*}

\item[$2.$] There is an isomorphism
\begin{gather*}
H^1(|M|)\simeq\mathfrak{H}^1_N(|M|).
\end{gather*}

\item[$3.$] And also an orthogonal decomposition
\begin{gather*}
\Omega^2(|\partial M|)=\mathfrak{H}^2(|\partial M|)\oplus d\Omega^1(|\partial M|)\oplus d^\star\Omega^3(|\partial M|).
\end{gather*}
\end{enumerate}
\end{lma}

The proof follows from restating arguments used in~\cite[Theorems~7 and~8]{AGO}.

Def\/ine
\begin{gather*}
{\Phi}_{A_M}:= L_{M}\cap\Omega^1_N(|M|)\subset L_{M}.
\end{gather*}
Notice that $r_M({\Phi}_{A_M})\subset {L_{\partial M}}$.
If $\varphi\in\Omega^k(|\partial M_\varepsilon|)$ satisf\/ies the Neumann condition $X^*_{\partial M}(\star\varphi)=0$,
then it also satisf\/ies $X^*_{{\partial M}}(\iota_{\partial_\tau}\varphi)=0$ with
$\iota_{\partial_\tau}\varphi\in\Omega^{k-1}(|\partial M_\varepsilon|)$.

Let us consider coclosed f\/ields which, according to Lemma~\ref{lma:orthogonaldecompostion}, have an
orthogonal decomposition of the axial gauge f\/ixing space of solutions
\begin{gather*}
\Phi_{A_M}= \big\{\overline{\varphi}_\mathfrak{h}+d^\star\alpha \in L_M \,|\,
\overline{\varphi}_\mathfrak{h}\in\mathfrak{H}_N^1(|M|), \, \alpha\in\Omega^2_N(|M|), \, d^\star dd^\star\alpha=0=dd^\star
d\alpha \big\}.
\end{gather*}
If $\overline{\varphi}\in \Phi_{A_M}$, then $\overline{\varphi}$ satisf\/ies the equation $d^\star\overline{\varphi}=0$,
it also satisf\/ies the Euler--Lagrange equation $d^\star d\overline{\varphi}=0$, since $d^\star dd^\star\alpha=0=dd^\star
d\alpha$.
This space
\begin{gather*}
\Phi_{A_M}\subset \mathfrak{H}_N^1(|M|)\oplus d^\star\Omega^2_N(|M|),
\end{gather*}
constitutes the orthogonal projection of the space of solutions $L_M$, according to the decomposition
\begin{gather*}
\Omega^1(|M|)=\mathfrak{H}^1_N(|M|)\oplus d^\star\Omega^2_N(|M|)\oplus
d\Omega^0_D(|M|)\oplus\big(\mathfrak{H}^1(|M|)\cap d\Omega^0(|M|)\big).
\end{gather*}
From this orthogonal decomposition it can be
shown that every solution $\varphi\in L_M$ can be transformed, modulo the bulk gauge transformation,
\begin{gather*}
\varphi\mapsto \overline{\varphi}=\varphi+df
\end{gather*}
onto a~f\/ield belonging to the space~$\Phi_{A_M}$.
Thus the following statement can be proven.

\begin{lma}\label{lma:Phi_M}\qquad
\begin{enumerate}\itemsep=0pt

\item[$1.$] Every $G_M^0$-orbit intersects $\Phi_{A_M}\subset L_M$ in exactly one point, i.e., for every $\varphi\in L_M$ there
exists $f\in\Omega^0(|M|)$, such that $\overline{\varphi}=\varphi+df\in\Phi_{A_M}$.

\item[$2.$] ${r}_M\colon \Phi_{A_M}\rightarrow L_{\tilde{M}}\cap \Phi_{{A}_{\partial M}}$ is a~linear surjection.
\end{enumerate}
\end{lma}

\begin{cor}
\label{pro:Lagrangianinclusion}
$L_{\tilde{M}}/G^0_{\partial M}=r_M(\Phi_{A_M})/G^0_{\partial M}\subset L_{\partial M}/G_{\partial M}^0$.
\end{cor}

Consider the identif\/ication $\mathfrak{H}^1_N(|M|)\simeq\mathfrak{H}^1(|\partial M|)$, and $d^\star_N\Omega^2(|M|)\simeq
d^{\star_{\partial M}}\Omega^2(|\partial M|)$.
We have the following statement.

\begin{lma}\label{lma:id}\quad
\begin{enumerate}\itemsep=0pt
\item[$1.$] There is a~well defined restriction map
\begin{gather*}
X_{\partial M}^*\colon \ \Phi_{A_M}\rightarrow \mathfrak{H}^1(|\partial M|)\oplus d^{\star_{\partial M}}\Omega^2(|\partial M|)
\end{gather*}

\item[$2.$] If we adopt the identification given in Lemma~{\rm \ref{tma:1}},
\begin{gather*}
L_{\partial M}/G_{\partial M}^0\simeq T\big(\mathfrak{H}^1(|\partial M|)\big)\times T\big(d^{\star_{\partial
M}}\Omega^2(|\partial M|)\big),
\end{gather*}
then the map ${r}_M$ coincides with the first jet of the pullback, i.e., we have a~commutative diagram of linear mappings
\begin{gather*}
\xymatrix{
\Phi_{A_M}\ar@{^{(}->}[d]\ar[r]^{r_M}&
L_{{\partial{ M}}} /G^0_{\partial M}\ar@{<->}[d]
\\
\mathfrak{H}_N^1(|M|)\oplus d^\star\Omega^2_N(|M|)
\ar[r]_{j^1X_{\partial M}^*}&
T\big(\mathfrak{H}^1(|\partial M|)\big)\times T\big(d^{\star_{\partial M}}\Omega^2(|\partial M|)\big)
}
\end{gather*}
\end{enumerate}
\end{lma}

As a~matter of fact de Rham theorems also apply in stratif\/ied spaces.
There is an isomorphism for the singular cohomology
\begin{gather*}
H^1(|\partial M|,\mathbb{R})\simeq \mathfrak{H}^1_{dR}(|\partial M|)
\end{gather*}
only in the case when the stratif\/ication is complete, see def\/inition in~\cite[Theorem~18]{AGO}.
This occurs for instance in the case where the faces $\Sigma^i\subset\partial M$ are homeomorphic to balls.
In general there is just a~morphism $H^1_{dR}(|\partial M|)\rightarrow \mathfrak{H}^1(|\partial M|)$.

Now we are in a~position to prove the result that encodes dynamics in a~Lagrangian context.
We prove that the image of solutions modulo gauge onto the space of boundary conditions modulo gauge, stated in
Proposition~\ref{pro:Lagrangianinclusion}, is in fact a~Lagrangian space.
The following statement completes the dynamical picture described in Axiom~\ref{ax:9}.

\begin{tma}
\label{tma:Lagrangian}
Let $L_{\tilde{M}}=r_M(L_M)$ be the boundary conditions that can be extended to solutions in the interior~$L_M$.
Then for the symplectic vector space $L_{\partial M}/G^0_{\partial M}$
\begin{enumerate}\itemsep=0pt
\item[$1.$] $L_{\tilde{M}}/G^0_{\partial M}$ is an isotropic subspace.
\item[$2.$] $L_{\tilde{M}}/G^0_{\partial M}$ is a~coisotropic subspace.
\end{enumerate}
In other words $L_{\tilde{M}}\cap \Phi_{A_{\partial M}}$ is a~Lagrangian subspace of the symplectic space
$\Phi_{A_{\partial M}}$.
\end{tma}

As we mentioned in the introduction for $L_{\tilde{M}}/G^0_{\partial M}$ isotropy is always true, see~\cite{KT}.
For the sake of completeness we give a~proof that is a~straightforward calculation.
Take $\varphi,\varphi'\in L_M$ and consider its image
\begin{gather*}
\big(j^1X^*_{\partial M}\big)(\varphi)=\big(X^*_{\partial M}\overline{\varphi},X^*_{\partial
M} (\mathcal{L}_{\partial_\tau}\overline{\varphi} )\big)=\big(\phi^\varphi,\dot{\phi}^{\varphi}\big)\in
L_{\partial M}\cap\Phi_{A_{\partial M}}
\end{gather*}
where $\overline{\varphi}$ was def\/ined in Lemma~\ref{pro:1}, and where $d^{\star_{\partial
M}}\phi^\varphi=d^{\star_{\partial M}}\dot{\phi}^{\varphi}=0$.
We also consider $\overline{\varphi}\in \Phi_{A_M}$.
Then
\begin{gather*}
\omega_{\partial M}\big(\overline{\varphi},\overline{\varphi}{}'\big)= \frac{1}{2}\int_{\partial
M}\phi^{{\varphi}}\wedge\star_{\partial M}\dot{\phi}^{\varphi'}- \phi^{\varphi'}\wedge\star_{\partial
M}\dot{\phi}^{\varphi}
\\
\hphantom{\omega_{\partial M}\big(\overline{\varphi},\overline{\varphi}{}'\big)}{}
=
\frac{1}{2}\int_{\partial M}X^*_{\partial M}(\overline{\varphi})\wedge\star_{\partial M}X_{\partial
M}^*\big(\mathcal{L}_{\partial_\tau}\overline{\varphi}{}'\big)- X^*_{\partial M}\big(\overline{\varphi}{}'\big)\wedge\star_{\partial
M}X_{\partial M}^*(\mathcal{L}_{\partial_\tau}\overline{\varphi}).
\end{gather*}
From a~property shown in Lemma~\ref{lma:embeddingndim} we have that the last expression equals
\begin{gather*}
\frac{1}{2}\int_{\partial M}X^*_{\partial M}\big(\overline{\varphi}\wedge\star
d\overline{\varphi}{}'-\overline{\varphi}{}'\wedge\star d\overline{\varphi}\big).
\end{gather*}
Recall hat $\varphi$, $\varphi'$ are global solutions in the interior $d^\star d\varphi=0=d^\star d\varphi'$, hence~by
applying Stokes' theorem we have
\begin{gather*}
\omega_{\partial M}\big(\overline{\varphi},\overline{\varphi'}\big) =\int_Md\overline{\varphi}\wedge\star
d\overline{\varphi'}- d\overline{\varphi'}\wedge\star d\overline{\varphi}=0.
\end{gather*}

\begin{proof}[Proof for coisotropic embedding] \sloppy Take $\overline{\varphi}\in\Phi_{A_M}$, as indicated in Lemma~\ref{lma:Phi_M}, take
$(\phi^{\varphi},\dot{\phi}^{\varphi}) :=r_M(\overline{\varphi})$ and suppose that
$\omega_{\partial M}(\overline{\varphi},\overline{\varphi}{}^\varepsilon)=0$ for every $\overline{\varphi}{}^\varepsilon\in
L_{\partial M_\varepsilon}$ with $(\phi',\dot{\phi}')\in L_{\partial M}$ corresponding to
$\overline{\varphi}{}^\varepsilon$ with $\iota_{\partial_\tau}\overline{\varphi}{}^{\varepsilon}$.
Recall that $d^{\star_{\partial M}}\phi'=0=d^{\star_{\partial M}}\dot{\phi'}$.
Then thanks to the representative~\eqref{eqn:Xwedge} we have
\begin{gather}
\label{eqn:omega=0}
\int_{\partial M}\phi^{\varphi}\wedge\star_{\partial M}\dot{\phi}'=\int_{\partial M}\phi'\wedge\star_{\partial
M}\dot{\phi}^{\varphi},
\qquad
\forall\, \big(\phi',\dot{\phi}'\big)\in L_{\partial M}.
\end{gather}

According to the orthogonal decomposition described in Lemma~\ref{lma:orthogonaldecompostion}, we have
\begin{gather*}
\phi^\varphi=\phi_\mathfrak{h}+d^{\star_{\partial M}}X^*_{\partial M}(\alpha),
\qquad
\dot{\phi}^{{\varphi}}=\dot{\phi}_\mathfrak{h}+d^{\star_{\partial M}}X^*_{\partial M}(\dot{\alpha}),
\end{gather*}
where
\begin{gather*}
{\phi}_\mathfrak{h},\dot{\phi}_\mathfrak{h}\in\mathfrak{H}^1(|\partial M|),
\qquad
\alpha,\dot{\alpha}\in\Omega^2(|\partial M_\varepsilon|), \qquad \dot{\alpha}=\mathcal{L}_{\partial_\tau}\alpha,\qquad
d^*dd^*\alpha=0
\end{gather*}
Hence equation~\eqref{eqn:omega=0} implies $\forall\, (\phi',\dot{\phi}')\in L_{\partial M}$
\begin{gather}
\int_{\partial M}\phi_\mathfrak{h}\wedge\star_{\partial M}\dot{\phi}'+ \int_{\partial M}d^{\star_{\partial
M}}X^*_{\partial M}(\alpha)\wedge\star_{\partial M}\dot{\phi}'\nonumber
\\
\qquad {}=\int_{\partial M}\phi'\wedge\star_{\partial M}\dot{\phi}_{\mathfrak{h}}+ \int_{\partial M}\phi'\wedge\star_{\partial
M}d^{\star_{\partial M}}X^*_{\partial M}(\dot{\alpha}).\label{eqn:omega=0_A}
\end{gather}
We calculate in more detail the f\/irst summand of the r.h.s.\
of equation~\eqref{eqn:omega=0_A}.
According to Lemma~\ref{lma:id}, $\dot{\phi}_\mathfrak{h}=X^*_{\partial M}(\mathcal{L}_{\partial
\tau}\overline{\varphi}_\mathfrak{h})$, where we consider the orthogonal decomposition
\begin{gather*}
\overline{\varphi}=\overline{\varphi}_\mathfrak{h}+d^*{\alpha}\in\Phi_{A_M}\subset \mathfrak{H}^1_N(|M|)\oplus
d^*\Omega^2_N(|M|)
\end{gather*}
with $\overline{\varphi}_\mathfrak{h}\in\mathfrak{H}_N^1(|M|)$, ${\alpha}\in\Omega^2_N(|M|)$.
Hence{\samepage
\begin{gather*}
\int_{\partial M}\phi'\wedge\star_{\partial M}\dot{\phi}_{\mathfrak{h}}= \int_{\partial M}\phi'\wedge\star_{\partial
M}X^*_{\partial M}\big(\mathcal{L}_{\partial \tau}\overline{\varphi}_\mathfrak{h}\big)= \int_{\partial
M}\phi'\wedge\star_{\partial M}X^*_{\partial M}\big(d\overline{\varphi}_\mathfrak{h}\big)= 0.
\end{gather*}
In the last line we have used the properties described for $X_{\partial M}^*$ given in Lemma~\ref{lma:embeddingndim} and
$d\overline{\varphi}_{\mathfrak{h}}=0$.}

Now consider the f\/irst summand of the l.h.s.
of equation~\eqref{eqn:omega=0_A}, the extension
$\tilde{\varphi}:=\psi\cdot\overline{\varphi^\varepsilon}\in\Omega^1_N(|M|)$ of $\overline{\varphi^\varepsilon}\in
\Omega^1	(|\partial M_\varepsilon|)$, given by a~partition of unity
\begin{gather}
\label{eqn:psi}
\psi\colon \ M\rightarrow [0,1], \qquad \text{such that} \quad \partial M=\psi^{-1}(1).
\end{gather}
Then
\begin{gather*}
\int_{\partial M}\phi_\mathfrak{h}\wedge\star_{\partial M}\dot{\phi}'= \int_{\partial
M}\phi_\mathfrak{h}\wedge\star_{\partial M}X^*_{\partial M} (\mathcal{L}_{\partial_\tau}{\tilde{\varphi}} )=
\int_{\partial M}\phi_\mathfrak{h}\wedge X^*_{\partial M} (\star d{\tilde{\varphi}} ).
\end{gather*}
Furthermore, Lemma~\ref{lma:id} claims that there exists $\overline{\varphi}_\mathfrak{h}\in \mathfrak{H}^1_N(|M|)$ such
that $\phi_\mathfrak{h}=X^*_{\partial M}(\overline{\varphi}_\mathfrak{h})$.
Therefore by Stokes' theorem
\begin{gather*}
\int_{\partial M}\phi_\mathfrak{h}\wedge\star_{\partial M}\dot{\phi}'= \int_{\partial M}X^*_{\partial
M} (\overline{\varphi}_\mathfrak{h}\wedge\star d{\tilde{\varphi}} )=
\int_{M}d (\overline{\varphi}_\mathfrak{h}\wedge\star d{\tilde{\varphi}} )=0.
\end{gather*}
Therefore equation~\eqref{eqn:omega=0_A} yields
\begin{gather}
\label{eqn:omega=0_B}
\int_{\partial M}d^{\star_{\partial M}}X_{\partial M}^*(\alpha)\wedge\star_{\partial M}\dot{\phi}'= \int_{\partial
M}\phi'\wedge\star_{\partial M}d^{\star_{\partial M}}X_{\partial M}^*(\dot{\alpha}).
\end{gather}
According to the isomorphism $H^1(|M|)\simeq\mathfrak{H}^1_N(|M|)$ in Lemma~\ref{lma:orthogonaldecompostion},
$[\tilde{\varphi}]\in H^1(|M|)$ corresponds to a~harmonic f\/ield
\begin{gather}
\label{eqn:phi_h}
[\tilde{\varphi}]\leftrightarrow\overline{\varphi}'_\mathfrak{h}\in\mathfrak{H}^1_N(|M|),
\end{gather}
and this in turn corresponds to the harmonic component $\phi_\mathfrak{h}'\in \mathfrak{H}^1(|{\partial M}|)$ of
\begin{gather*}
X^*_{\partial M} (\overline{\varphi'} )=\phi'=\phi'_\mathfrak{h}+d^{\star_{\partial M}}\beta
\end{gather*}
with $\beta\in\Omega^2(|\partial M|)$, as is stated in Lemma~\ref{lma:id}.
Thus, for the r.h.s.\
of equation~\eqref{eqn:omega=0_B} we have
\begin{gather*}
\int_{\partial M}\phi'_\mathfrak{h}\wedge\star_{\partial M}d^{\star_{\partial M}}X_{\partial M}^*(\dot{\alpha})+
\int_{\partial M}d^{\star_{\partial M}}\beta\wedge\star_{\partial M}d^{\star_{\partial M}}X_{\partial
M}^*(\dot{\alpha}).
\end{gather*}
Notice that $\partial\partial M=0$, therefore the last expression equals
\begin{gather*}
\int_{\partial M}d\phi'_\mathfrak{h}\wedge\star_{\partial M}X_{\partial M}^*(\dot{\alpha})+ \int_{\partial
M}d^{\star_{\partial M}}\beta\wedge\star_{\partial M}d^{\star_{\partial M}}X_{\partial M}^*(\dot{\alpha})
\\
\qquad {} = \int_{\partial M}d^{\star_{\partial M}}\beta\wedge\star_{\partial M}d^{\star_{\partial M}}X_{\partial
M}^*(\dot{\alpha}).
\end{gather*}
Similarly for $\dot{\phi}'=\dot{\phi}_\mathfrak{h}+d^{\star_{\partial M}}\dot{\beta}$ with
$\dot{\phi}_\mathfrak{h}\in\mathfrak{H}^1(|\partial M|)$, $\dot{\beta}\in\Omega^2(|\partial M|)$ and therefore the l.h.s.
of equation~\eqref{eqn:omega=0_B} equals $\int_{\partial M}d^{\star_{\partial M}}\dot{\beta}\wedge\star_{\partial
M}d^{\star_{\partial M}}X_{\partial M}^*({\alpha})$.
\begin{gather}
\label{eqn:beta}
\int_{\partial M}\beta\wedge\star_{\partial M}dd^{\star_{\partial M}}X_{\partial M}^*(\dot{\alpha})= \int_{\partial
M}\dot{\beta}\wedge\star_{\partial M}dd^{\star_{\partial M}}X_{\partial M}^*({\alpha}).
\end{gather}
Finally this eqution describes a~condition on pairs $\beta,\dot{\beta}\in\Omega^2(|\partial M|)$, $\forall\,
\alpha,\dot{\alpha}\in\Omega^2_N(|\partial M_\varepsilon|)\subset \Omega_N^2(|M|)$. Again by Stokes' theorem applied to
the r.h.s.\
of the previous expression~\eqref{eqn:beta} we have
\begin{gather}
\label{eqn:stokes}
\int_{\partial M}\beta\wedge\star_{\partial M}dd^{\star_{\partial M}}X_{\partial M}^*(\dot{\alpha})=
\int_{M}d\tilde{\dot{\beta}}\wedge\star d d^{\star}({\alpha}), \qquad \forall\, \alpha\in \Omega^2_N(|M|),
\end{gather}
where $\tilde{\dot{\beta}}=\psi\cdot\dot{\beta}\in\Omega_N^2(|M|)$ is an extension of a~$2$-form in the cylinder
$\dot{\beta}\in \Omega^2(|\partial M|)$ to the interior of~$M$, given by a~partition of unity~$\psi$,~\eqref{eqn:psi}.

Recall that since $\overline{\varphi}\in\Phi_{A_M}$, then $d^\star dd^\star\alpha=0$.
From the orthogonal decomposition
\begin{gather*}
\Omega^3(|M|)=d\Omega^2_D(|M|)\oplus\mathfrak{H}^3_N(|M|)\oplus\big(\mathfrak{H}^3(|M|)\cap d\Omega^2(|M|)\big)
\oplus d^\star\Omega_N^4(|N|)
\end{gather*}
we have $dd^\star\alpha\in \mathfrak{H}^3(|M|)\cap d\Omega^2(|M|)$.
By the non-degeneracy of the Hodge inner product in~$M$, there is a~well def\/ined exact harmonic f\/ield
$d\check{\beta}\in\mathfrak{H}^3(|M|)\cap d\Omega^2$, that is the projection of $d\tilde{\dot{\beta}}$, such that
$d^\star d\check{\beta}=0$, and the r.h.s.\
of~\eqref{eqn:stokes} reads as
\begin{gather*}
\int_{M}d\check{{\beta}}\wedge\star d d^{\star}({\alpha}),
\end{gather*}
therefore
\begin{gather}
\label{eqn:beta_t}
\int_{\partial M}\beta\wedge\star_{\partial M}dd^{\star_{\partial M}}X_{\partial M}^*(\dot{\alpha}) =\int_{\partial
M}X^*_{\partial M}\left(\check{{\beta}}\right)\wedge\star_{\partial M} d d^{\star_{\partial M}}X^*_{\partial
M}({\alpha}).
\end{gather}
On the other hand consider the l.h.s.\
of~\eqref{eqn:stokes}.
Recall that $\beta\in\Omega^2(|\partial M|)=\mathfrak{H}^2(|\partial M|)\oplus d\Omega^1(|\partial M|)\oplus
d^*\Omega^3(|\partial M|)$, in fact we can take
\begin{gather*}
\beta=\beta_\mathfrak{h}+d\gamma\in\mathfrak{H}^2(|\partial M|)\oplus d\Omega^1(|\partial M|).
\end{gather*}
Consider the extension $\tilde{\beta}:=\psi\beta$,
\begin{gather*}
\tilde{\beta}=\tilde{\beta}_\mathfrak{h}+d(\check{\gamma}+\gamma_D)+d^*\theta\in\Omega^2(|M|)
\\
\hphantom{\tilde{\beta}}{}=
\mathfrak{H}^2_N(|M|)\oplus \big(\mathfrak{H}^2(|M|)\cap d\Omega^1(|M|)\big)\oplus d\Omega_D^1(|M|)\oplus
d^*\Omega_N^3(|M|).
\end{gather*}
If we take the orthogonal projection of $\tilde{\beta}$,
\begin{gather}
\label{eqn:netahat}
\hat{\beta}:=\tilde{\beta}_\mathfrak{h}+d\hat{\gamma}\in\mathfrak{H}^2_N(|M|)\oplus \big(\mathfrak{H}^2(|M|)\cap
d\Omega^1(|M|)\big)\oplus d\Omega^1_D(|M|), \hat{\gamma}=\check{\gamma}+\gamma_D,
\end{gather}
then $X^*_{\partial M}(\tilde{\beta}_\mathfrak{h})=\beta_\mathfrak{h}$ and $X_{\partial
M}^*(\hat{\gamma})=\gamma$. Also for $3$-forms as arguments of $dd^{\star_\partial M}$ we have the functionals
\begin{gather*}
\int_{\partial M}X_{\partial M}^* \big(\tilde{\beta} \big)\wedge\star_{\partial M}dd^{\star_{\partial
M}}\cdot=\int_{\partial M}X_{\partial M}^*\big(\tilde{\beta}_\mathfrak{h}+d\hat{\gamma}\big)\wedge\star_{\partial
M}dd^{\star_{\partial M}}\cdot,
\\
\int_{\partial M}({\beta}_\mathfrak{h}+d{\gamma})\wedge\star_{\partial M}dd^{\star_{\partial M}}\cdot= \int_{\partial
M}X_{\partial M}^*\big(\hat{\beta}\big)\wedge\star_{\partial M}dd^{\star_{\partial M}}\cdot.
\end{gather*}
If we look more carefully the l.h.s.\
of expression~\eqref{eqn:beta_t}, then
\begin{gather*}
\int_{\partial M}X_{\partial M}^*(\hat{\beta})\wedge\star_{\partial M}dd^{\star_{\partial M}}X_{\partial
M}^* (\mathcal{L}_{\partial_\tau}(\alpha) )= \int_{\partial M}X_{\partial M}^*(\hat{\beta})\wedge X_{\partial
M}^*\big(\mathcal{L}_{\partial_\tau}\star (dd^{\star}\alpha)\big)
\\
\qquad {}=\int_{\partial M}X_{\partial M}^*\mathcal{L}_{\partial_\tau}\big(\hat{\beta}\wedge\star (dd^{\star}\alpha)\big)+
\int_{\partial M}X_{\partial M}^* \big(\mathcal{L}_{\partial_\tau}\hat{\beta}\big)\wedge X_{\partial M}^*\big(\star
(dd^{\star}\alpha)\big)
\\
\qquad {}=\mathcal{L}_{\partial_\tau}\left(\int_{M}d\big(\hat{\beta}\wedge\star (dd^{\star}\alpha)\big)\right)+ \int_{\partial
M}X_{\partial M}^*\big(\mathcal{L}_{\partial_\tau}\hat{\beta}\big)\wedge X_{\partial M}^*\big(\star
(dd^{\star}\alpha)\big)
\\
\qquad {}=
\int_{\partial M}\dot{\beta}\wedge\star_{\partial M}dd^{\star_{\partial M}}X_{\partial M}^*({\alpha}),
\end{gather*}
where in the last equality we used $X^*_{\partial M}(\mathcal{L}_{\partial_\tau}\hat{\beta})=X^*_{\partial
M}(\mathcal{L}_{\partial_\tau}\beta)=\dot{\beta}$ and that
$\mathcal{L}_{\partial_\tau}\int_M\hat{\beta}\wedge\star dd^\star\alpha=0$.
Hence
\begin{gather*}
\int_{\partial M}X_{\partial M}^*(\hat{\beta})\wedge\star_{\partial M} dd^{\star_{\partial M}}X_{\partial
M}^*\left(\mathcal{L}_{\partial_\tau}(\alpha)\right)=		\int_{\partial M}\dot{\beta}\wedge\star_{\partial
M}dd^{\star_{\partial M}} X_{\partial M}^*({\alpha}).
\end{gather*}
Looking back again at expression~\eqref{eqn:beta_t} and~\eqref{eqn:beta} we have
\begin{gather*}
\int_{\partial M}\dot{{\beta}}\wedge\star_{\partial M} d d^{\star_{\partial M}} X^*_{\partial M}({\alpha})=
\int_{\partial M}\beta\wedge\star_{\partial M} dd^{\star_{\partial M}}X^*_{\partial M}(\dot {\alpha})
\\
\hphantom{\int_{\partial M}\dot{{\beta}}\wedge\star_{\partial M} d d^{\star_{\partial M}} X^*_{\partial M}({\alpha})}{}
=\int_{\partial M}X^*_{\partial M}\big(\check{{\beta}}\big)\wedge\star_{\partial M} d d^{\star_{\partial
M}}X^*_{\partial M}({\alpha}).
\end{gather*}
Hence for every $\alpha $ we have
\begin{gather*}
\int_{\partial M}X_{\partial M}^*\big(\mathcal{L}_{\partial_\tau}\hat{\beta}\big)\wedge X_{\partial M}^*\big(\star
(dd^{\star}\alpha)\big)= \int_{\partial M}X^*_{\partial M}\big(\check{{\beta}}\big)\wedge\star_{\partial M} d
d^{\star_{\partial M}}X^*_{\partial M}({\alpha}).
\end{gather*}
This implies that $X_{\partial M}^*(\mathcal{L}_{\partial_\tau}\hat{\beta})=X^*_{\partial
M}(\check{{\beta}})$.

Finally we can extend the solution $\overline{\varphi^\varepsilon}$ in the cylinder $\partial M_{\varepsilon}$ to
a~solution in the interior~$M$, by means of
\begin{gather*}
\overline{\varphi}':=\overline{\varphi}_\mathfrak{h}'+d^\star\big(\hat{\beta}\big),
\end{gather*}
where $\overline{\varphi}_\mathfrak{h}'$ was def\/ined in~\eqref{eqn:phi_h} and $\hat{\beta}$ is def\/ined
in~\eqref{eqn:netahat}.
Notice that $d^*dd^*\hat{\beta}=d^*dd^*d\gamma=0$, therefore $\overline{\varphi}'\in\Phi_{A_M}$.
Furthermore,
\begin{gather*}
\phi'=X^*_{\partial M}\big(\overline{\varphi}'\big),
\qquad
\dot{\phi}'=X^*_{\partial M}\big(\mathcal{L}_{\partial_\tau}\overline{\varphi}'\big).\tag*{\qed}
\end{gather*}
\renewcommand{\qed}{}
\end{proof}

This f\/inishes the proof of the validity of Axiom~\ref{ax:9}.

{\sloppy As we mentioned in Subsection~\ref{subsec:locality}, locality follows for f\/ields and actions, in particular \mbox{Axiom~\ref{ax:11}}
hold.
The gluing Axiom~\ref{ax:12} also follows from locality arguments.
This completes the dynamical description for this gauge f\/ield theory.

}

Thus abelian theory is fully constructed within this axiomatic framework.

\section{Example: 2-dimensional case}
\label{sec:4}

For a~better understanding of our model, we review our constructions in a~more down to earth example, namely the
$2$-dimensional case.
We provide this presentation as a~comparison tool with some quantizations of two-dimensional theories, see for
instance~\cite{DH,La, Wi}.
This also suggest the steps that are necessary in quantization for general dimensions in further research.

Recall that we are supposing that we have a~trivial gauge principal bundles on a~compact surface~$M$, with structure
group $G=U(1)$.
The following lemma will lead to a~description of the presymplectic structure $\tilde{\omega}_\Sigma$, on
$\tilde{A}_\Sigma$, for a~proof see~\cite{Il}.
Lemma~\ref{lma:embeddingndim} in this case can be simplif\/ied as the following statement.

\begin{lma}[Fermi] Given a~cylinder $\Sigma\times[0,1]$, there exists an embedding
\begin{gather*}
X\colon \ \Sigma\times[0,\varepsilon]\rightarrow M
\end{gather*}
of the cylinder $\Sigma\times [0,\varepsilon]$ into a~tubular neighborhood $\Sigma_\varepsilon$ of~$\Sigma$, such that
if $(s,\tau)$ are local coordinates, then $\partial/\partial s$, $\partial /\partial \tau$ are orthonormal
vector fields along~$\Sigma$.
Here~$s$ corresponds to arc length along~$\Sigma$ with respect to the Riemannian metric~$h$ on~$M$.
Furthermore $h|_\Sigma$ is locally described as the identity matrix.
\end{lma}

The presymplectic structure can be written by using these local coordinate as in~\eqref{eqn:presymplectic},
\begin{gather*}
\tilde{\omega}_\Sigma\big(\tilde{\eta},\tilde{\xi}\big)=
\frac{1}{2}\int_\Sigma\big[\eta^s\big(\partial_s\xi^\tau-\partial_\tau\xi^s\big)-\xi^s\big(\partial_s\eta^\tau-\partial\tau\eta^s\big)\big]ds,
\end{gather*}
where $X^*_\Sigma(\eta)=\eta^sds+\eta^\tau d\tau, X^*_\Sigma(\xi)=\xi^sds+\xi^\tau d\tau$ are $1$-forms corresponding to
solutions in the cylinder, i.e., $\xi,\eta\in\Omega^1(\Sigma_\varepsilon)$ satisfying  Euler--Lagrange equations.
We can also describe the gauge group $G_\Sigma$ on $A_\Sigma$, by considering the action of the identity component gauge
group of germs: $\tilde{G}_\Sigma^0:=\varinjlim G^0_{\Sigma_\varepsilon}$. Here
${G}_{\Sigma_\varepsilon}^0:=\Omega^0(\Sigma_\varepsilon)/\mathbb{R}^{b_0}$ is acting by translations $\eta \mapsto \eta
+ df$ and inducing the corresponding action $\tilde{\eta}\mapsto\tilde{\eta}+d\tilde{f}$ on germs $\tilde{\eta}\in
\tilde{L}_\Sigma$.

The degeneracy subspace of the symplectic form is
\begin{gather*}
{K}_{\omega_\Sigma}:=\big\{\tilde{\eta}\in \tilde{L}_\Sigma \,|\, {\eta}=\partial_\tau f d\tau, \, \partial_sf(s,0)=0,\,
f\in		\Omega^0(\Sigma_\varepsilon)\big\}.
\end{gather*}
From this very def\/inition we have that the degeneracy gauge symmetry group $K_{\omega_\Sigma}$ is a~(normal) subgroup of
the abelian group~$\tilde{G}_\Sigma^0$.

By considering an axial gauge f\/ixing, as in~\eqref{eqn:gaugefinxing}, let
\begin{gather*}
\Phi_{\tilde{A}_\Sigma}:=\big\{\overline{\eta}\in \tilde{L}_\Sigma \,|\, \iota_{\partial_\tau}\overline{\eta}=0\big\}
\end{gather*}
be a~subspace of $\tilde{L}_\Sigma$.
As we did in Lemma~\ref{pro:1} we have that every ${K}_{\omega_\Sigma}$-orbit in $\tilde{L}_\Sigma$ intersects in just
one point the subspace $\Phi_{\tilde{A}_\Sigma}$.
The presymplectic form $\tilde{\omega}_\Sigma$ restricted to the subspace $\Phi_{\tilde{A}_\Sigma}$ may be written as{\samepage
\begin{gather}
\label{eqn:omega-2}
\tilde{\omega}_\Sigma(\tilde{\eta},\tilde{\xi})=\frac{1}{2}\int_\Sigma\big[{-}\eta^s\partial_\tau\xi^s+\xi^s\partial_\tau\eta^s\big]ds,
\qquad
\tilde{\xi},\tilde{\eta}\in \tilde{L}_{\Sigma}.
\end{gather}
Hence $\tilde{\omega}_\Sigma$ is non-degenerate when we restrict it to the subspace $\Phi_{\tilde{A}_\Sigma}\subset
\tilde{L}_\Sigma$.}

Let $\omega_\Sigma$ the corresponding \emph{symplectic} structure on $A_\Sigma$ induced by the restriction of
$\tilde{\omega}_\Sigma$ to the subspace $\Phi_{\tilde{A}_\Sigma}\subset \tilde{L}_\Sigma$.

Hypersurfaces are $\Sigma:=\Sigma^1\cup\dots\cup \Sigma^m\subset \partial M$.
In the smooth case $\Sigma^i$ are homeomorphic to~$S^1$.
In the case with corners, $\Sigma^i$ are intervals identif\/ied in some pairs by their boundaries.

Then there is a~linear map
\begin{gather*}
\Omega^1_{\partial M}\rightarrow \Omega^1\big(\Sigma^1\big)\oplus\dots\oplus\Omega^1\big(\Sigma^m\big),
\end{gather*}
where $\eta\mapsto\big(X_0^{\Sigma^1}\big)^*(\eta)\oplus\dots\oplus\big(X_0^{\Sigma^m}\big)^*(\eta)$.
This map induces $r_{\partial M;\Sigma}\colon L_{\partial M}\rightarrow=L_{\Sigma^1}\oplus\dots\oplus L_{\Sigma^m}$.
Furthermore, the chain decomposition $\int_{\partial M}\cdot =\int_{\Sigma^1}\cdot+\dots+\int_{\Sigma^m}\cdot$ induces
Axioms~\ref{ax:7}$'$ and~\ref{ax:7}.

Recall that here is a~map, $\tilde{r}_M\colon L_M\rightarrow \tilde{L}_{\partial M}$, coming from the restriction of the
solutions to germs on the boundary.
Composing with the quotient class map, we have a~map $r_M\colon L_M\rightarrow L_{\partial M}$.
Let $L_{\tilde{M}}$ be the image $r_M(L_M)$ under this map.

Our aim is to describe the image $L_{\tilde{M}}=r_M(L_M)\subset L_\Sigma$ of the space of solutions as a~Lag\-rangian
subspace modulo gauge.

Take $\varphi\in L_M$ so that $d^\star d\varphi=0$, then $d\varphi$ is constant a~scalar multiple of the~$h$-area
form~$\mu$, i.e., $d\varphi=\dot{c}_\varphi \mu$, for a~constant $\dot{c}_\varphi$.
Suppose that $\overline{\varphi}\in L_M$ is such that $\iota_{\partial_\tau}\overline{\varphi}=0$, then
$\overline{\varphi}^\tau=0$.
Hence $\partial_s\overline{\varphi}^\tau-\partial_\tau\overline{\varphi}^s=\dot{c}_\varphi$ is constant.
That is, $-\partial_\tau\overline{\varphi}^s=\dot{c}_\varphi$.
Therefore if $\phi:=r_M(\overline{\varphi})$, $\phi':=r_M(\overline{\varphi}')\in L_{\partial M}$, then by substituting
in~\eqref{eqn:omega-2} we obtain
\begin{gather*}
\omega_{\partial M}\big(\overline{\varphi},\overline{\varphi}'\big)= \int_{\partial M}\left(\varphi^s
\dot{c}_{\varphi'} - (\varphi')^s\dot{c}_\varphi\right) ds, \qquad \forall \, \overline{\varphi},\overline{\varphi}'\in
L_{\partial M}.
\end{gather*}
Recall that $\varphi,\varphi'\in L_{\partial M_\varepsilon}$.
By Stokes' theorem
\begin{gather}
\label{eqn:0}
\omega_{\partial M}\left(\overline{\varphi},\overline{\varphi}'\right)=
\dot{c}_{\varphi'}\int_Md\overline{\varphi}-\dot{c}_\varphi\int_M d\overline{\varphi}'=(\dot{c}_{\varphi'}
\dot{c}_\varphi-\dot{c}_\varphi \dot{c}_{\varphi'})\cdot\mathrm{area}(M)=0,
\end{gather}
where $\mathrm{area}(M):=\int_M\mu$.

We now consider the orbit space for gauge orbits.
We consider the unit component subgroup $G_{\Sigma}^0\trianglelefteq G_{\Sigma}$.
Recall the map~\eqref{eqn:A_Sigma}.
Take the gauge f\/ixing subspace
\begin{gather*}
\Phi_{A_\Sigma}:= \{(\eta_0,\dot{\eta}_0)\in A_\Sigma \,|\,
\partial_s\eta_0=0=\partial_s\dot{\eta}_0 \}=\big\{(c ds,\dot{c} ds)\in A_\Sigma \,|\,
(c,\dot{c})\in\mathbb{R}^2\big\}.
\end{gather*}
We can see the proof of Lemma~\ref{lma:intersection} for this context.
Let $(c ds,\dot{c} ds)$ be a~point in $\Phi_{A_\Sigma}\cong\mathbb{R}^2$.
Consider $X^*_\Sigma(\eta)=\eta^sds+\eta^\tau d\tau=\eta^sds$, a~local expression for a~solution $\eta\in
L_{\Sigma_\varepsilon}\cap \Phi_{\tilde{A}_{\Sigma}}$.
By considering a~gauge symmetry we can get an ODE for $f\colon  \Sigma_\varepsilon \rightarrow \mathbb{R}$,
\begin{gather}
\eta^s+\partial_s f=c,
\label{eqn:ODE1}
\\
\partial_\tau\eta^s+\partial_\tau\partial_s f=\dot{c}
\label{eqn:ODE2}
\end{gather}
Equation~\eqref{eqn:ODE2} can be solved for $g(s,\tau):=\partial_sf$, once we can f\/ix the boundary condition
$\partial_sf(s,0)=g(s,0)$.
This boundary condition in turn can be obtained by solving~\eqref{eqn:ODE1} in~$\Sigma$.
The holonomy along~$\Sigma$,
\begin{gather*}
\mathrm{hol}_\Sigma(\eta)=\exp \sqrt{-1}\int_\Sigma\eta\in G=U(1)
\end{gather*}
remains the same for~$c$ and for~$\eta$, furthermore since they are in the same component, $\int_\Sigma c ds$ equals
$\int_\Sigma \eta^sds$ mod $2\pi\mathbb{Z}$.
Here~$\eta$ belongs to the $G_\Sigma^0 $-orbit of~$c$, therefore there is a~homotopy between both evaluations.
Hence{\samepage
\begin{gather*}
c\cdot\mathrm{length}(\Sigma)=\int_\Sigma c ds=\int_\Sigma \eta^sds
\end{gather*}
this implies that equation~\eqref{eqn:ODE1} can be solved.}

Lemma~\ref{lma:Phi_M} is also satisf\/ied.
It follows that $L_{\tilde{M}}\cap \Phi_{A_{\partial M}}=r_M(\Phi_{A_M})$.
The isotropic embedding described in Theorem~\ref{tma:Lagrangian} is proved in~\eqref{eqn:0}.
The corresponding coisotropic embedding in the $2$-dimensional version goes as follows:

Take $\varphi\in\Phi_{A_M}$, $\phi=r_M(\varphi)$ and suppose that $\omega_{\partial M}(\phi,\phi')=0$ for every
$\varphi'\in L_{\partial M_\varepsilon}$, with $\phi'\in L_{\partial M}$ corresponding to $\varphi'$.
Then
\begin{gather*}
\dot{c}_\varphi\int_{\partial M}(\varphi')^s ds= \int_{\partial M}\left(\varphi^s\partial_\tau(\varphi')^s\right)ds.
\end{gather*}

Since $\varphi'$ is a~solution in a~tubular neighborhood $\partial M_\varepsilon$ then $\partial_\tau
(\varphi')^s|_\Sigma=\dot{c}_{\varphi'}$.
Thus
\begin{gather*}
\dot{c}_\varphi\int_{\partial M}(\varphi')^s ds= \dot{c}_{\varphi'}\int_{\partial M}\varphi^sds= \dot{c}_\varphi
\dot{c}_{\varphi'}\int_M\mu
\end{gather*}
therefore
\begin{gather*}
\int_{\partial M}\varphi'=\dot{c}_{\varphi'}\cdot\mathrm{area}(M).
\end{gather*}
We claim that this is a~suf\/f\/icient condition, so that $\varphi'\in L_{\partial M_\varepsilon}$ can be extended to the
interior of~$M$.
There exists a~solution $\check{\varphi}\in L_M$ such that $\varphi'=r_M(\check{\varphi})$.
This will be an exercise of calculus of dif\/ferential forms.

The f\/irst step is to construct an extension $\theta =\psi\varphi \in \Omega^1(M)$, where we take a~partition of
unity~$\psi$ whose value on $\partial M_\varepsilon $ is 1 and is $0$ outside an open neighborhood $V\subset M$ of
$\partial M_\varepsilon$.
We see that $\dot{c}_{\varphi'} d\theta$ is closed and also has the same relative de Rham cohomology class in
$H^2_{dR}(M,\partial M;\mathbb{R})$ as $\dot{c}_{\varphi'}\mu$.
Thus $\dot{c}_{\varphi'}(d\theta-\mu)=\dot{c}_{\varphi'} d\beta$, for a~$1$-form~$\beta$ such that $\beta|_{\partial
M}=0$.
Therefore we can def\/ine $\check{\varphi}:= \theta- \beta$ such that it is a~solution.
Therefore $d\check{\varphi}=\dot{c}_{\varphi'}\mu$ and it is also an extension,
$\check{\varphi}|_{M_\varepsilon}=\varphi'|_{M_\varepsilon}$.

This proves in the $2$-dimensional case the Lagrangian embedding of Theorem~\ref{tma:Lagrangian}.
Nevertheless we should notice that the proof of coisotropy is rather obvious in this case, since the reduced symplectic
space $A_\Sigma/G_\Sigma$ is f\/inite-dimensional.

Notice that in this case the bilinear form $[\cdot,\cdot]_\Sigma$ used in Axiom~\ref{ax:4}, corresponds to
\begin{gather*}
\big[\phi^\eta,\phi^\xi\big]_\Sigma:=-\int_\Sigma\big(\eta^s \partial_\tau\xi^s\big) ds.
\end{gather*}

Here $(c ds,\dot{c} ds)\in\Phi_{A_\Sigma}$ can be identif\/ied with $c+\sqrt{-1}\dot{c}\in\mathbb{C}$, provided with the
K\"ahler structure: $\mathrm{length}(\Sigma)\cdot dc\wedge d\dot{c}$.
The holonomy $\mathrm{hol}_\Sigma\colon \Omega^1(\Sigma)\rightarrow U(1)$ induces the derivative map $D
\mathrm{hol}_\Sigma\colon \Phi_{A_\Sigma}\rightarrow T U(1)$.
We have the following commutative diagram
\begin{gather*}
\xymatrix{
\Phi_{A_\Sigma}\ar@{-->}[r]_{D\,\mathrm{hol}_\Sigma}\ar@{<->}[d]& T U(1)\ar@{<->}[d]\\
\mathbb{C}\ar[r]^{\exp}&\mathbb{C}^\times
}
\end{gather*}

We can f\/inally def\/ine the reduced space as the topological cylinder
\begin{gather*}
{A_\Sigma}/G_\Sigma:=\Phi_{A_\Sigma}/\overline{G_\Sigma}\leftrightarrow \mathbb{C}^\times,
\end{gather*}
where $\overline{G_\Sigma}:=G_\Sigma/G^0_{\Sigma}\simeq\mathbb{Z}$.
We can get the symplectic structure $\overline{\omega_\Sigma}$ on $A_\Sigma/G_\Sigma$.
This $\overline{\omega_\Sigma}$ is $\mathrm{length}(\Sigma)$ times the area form on the cylinder $TU(1)$.
The \emph{reduced symplectic structure}: $\overline{\omega_\Sigma}$ on~$A_\Sigma/G_\Sigma$, is $\mathrm{length}(\Sigma)$
times the area form on the cylinder $TU(1)$.
For $\partial M=\Sigma^1\cup\dots\cup\Sigma^m$:
\begin{gather*}
A_M/G_M\rightarrow A_{\partial M}/G_{\partial M}=TU(1)\times\dots\times TU(1).
\end{gather*}
The space~$L_{M}$ has Lagrangian image which in each factor is the quotiented line:
\begin{gather*}
\big\{(c,\dot{c})\in\mathbb{R}^2 \,|\, c\cdot\mathrm{length}(\partial
M)=\dot{c}\cdot\mathrm{area}(M)\big\}/\mathbb{Z}\subset TU(1).
\end{gather*}

As a~consequence the map $A_M/G_M\rightarrow A_{\partial M}/G_{\partial M}$ does depend on global data of the metric,
such as $\mathrm{area}(M)$ and $\mathrm{length}(\partial M)$.
Recall that the same global dependence of dynamics holds for the quantum version, i.e., the quantum TQFT version of gauge
f\/ields.

Once we have completed reduction, the picture of quantization on this f\/inite-dimensional space can be specif\/ied,
cf.~\cite{DH,La,Wi}.
For a~complete description of quantization in $2$-dimensions in general non abelian case with corners see~\cite{O2}.

\section{Outlook: quantization in higher dimensions}

The geometric quantization program with corners will be treated elsewhere.
Once the reduction-quantization procedure is completed, the next task is the formulation of the quantization-reduction
process and the equivalence of both procedures.
See the discussion of these issues in dimension two for instance in~\cite{DH,La, Wi}.
In order to administer the geometric quantization program~\cite{Wo} for the reduced space we need to describe a~suitable
hermitian structure in~$\Phi_{A_\Sigma}$.
Another work in progress with more physical applications is the formulation corresponding to Lorentzian manifolds rather
than the Riemannian case.

\subsection*{Acknowledgements}

The author thanks R.~Oeckl for several discussions and encouragement for writing this note at CCM-UNAM.
This work was partially supported through a~CONACYT-M\'exico postdoctoral grant.
The author also thanks the referees for their comments and suggestions.

\pdfbookmark[1]{References}{ref}
\LastPageEnding


\begin{thebibliography}{99}
\footnotesize \itemsep=-0.5pt

\bibitem{AM}
Al-Zamil Q.S.A., Montaldi J., Witten--{H}odge theory for manifolds with
  boundary and equivariant coho\-mo\-lo\-gy, \href{http://dx.doi.org/10.1016/j.difgeo.2011.11.002}{\textit{Differential Geom. Appl.}}
  \textbf{30} (2012), 179--194, \href{http://arxiv.org/abs/1004.2687}{arXiv:1004.2687}.

\bibitem{AGO}
Ancona V., Gaveau B., Okada M., Harmonic forms and cohomology of singular
  stratif\/ied spaces, \href{http://dx.doi.org/10.1016/j.bulsci.2006.09.001}{\textit{Bull. Sci. Math.}} \textbf{131} (2007), 422--456.

\bibitem{ArM}
Arbieto A., Matheus C., A pasting lemma and some applications for conservative
  systems, \href{http://dx.doi.org/10.1017/S014338570700017X}{\textit{Ergodic Theory Dynam. Systems}} \textbf{27} (2007),
  1399--1417, \href{http://arxiv.org/abs/math.DS/0601433}{math.DS/0601433}.

\bibitem{At}
Atiyah M., Topological quantum f\/ield theories, \textit{Inst. Hautes \'Etudes
  Sci. Publ. Math.}  (1988), 175--186.

\bibitem{CMR}
Cattaneo A.S., Mnev P., Reshetikhin N., Classical and quantum {L}agrangian
  f\/ield theories with boundary, \textit{PoS Proc. Sci.}  (2011),
  PoS(CORFU2011), 044, 25~pages, \href{http://arxiv.org/abs/1207.0239}{arXiv:1207.0239}.

\bibitem{CMR1}
Cattaneo A.S., Mnev P., Reshetikhin N., Semiclassical quantization of classical
  f\/ield theories, in Mathematical Aspects of Quantum Field Theories, Editors
  D.~Calaque, T.~Strobl, \href{http://dx.doi.org/10.1007/978-3-319-09949-1_9}{\emph{Mathematical Physics Studies}}, Springer, Berlin, 2015,
  275--324, \href{http://arxiv.org/abs/1311.2490}{arXiv:1311.2490}.

\bibitem{DaM}
Dacorogna B., Moser J., On a partial dif\/ferential equation involving the
  {J}acobian determinant, \textit{Ann. Inst. H.~Poincar\'e Anal. Non
  Lin\'eaire} \textbf{7} (1990), 1--26.

\bibitem{DH}
Driver B.K., Hall B.C., Yang--{M}ills theory and the {S}egal--{B}argmann
  transform, \href{http://dx.doi.org/10.1007/s002200050555}{\textit{Comm. Math. Phys.}} \textbf{201} (1999), 249--290,
  \href{http://arxiv.org/abs/hep-th/9808193}{hep-th/9808193}.

\bibitem{Du}
Duf\/f G.F.D., Dif\/ferential forms in manifolds with boundary, \href{http://dx.doi.org/10.2307/1969770}{\textit{Ann. of
  Math.}} \textbf{56} (1952), 115--127.

\bibitem{DuS}
Duf\/f G.F.D., Spencer D.C., Harmonic tensors on {R}iemannian manifolds with
  boundary, \href{http://dx.doi.org/10.2307/1969771}{\textit{Ann. of Math.}} \textbf{56} (1952), 128--156.

\bibitem{Ge}
Gelca R., Topological quantum f\/ield theory with corners based on the {K}auf\/fman
  bracket, \href{http://dx.doi.org/10.1007/s000140050013}{\textit{Comment. Math. Helv.}} \textbf{72} (1997), 216--243,
  \href{http://arxiv.org/abs/q-alg/9603002}{q-alg/9603002}.

\bibitem{Ak}
Giachetta G., Mangiattori L., Sardanashvily G., Advanced classical f\/ield
  theory, \href{http://dx.doi.org/10.1142/9789812838964}{World Sci. Publ.}, Singapore, 2009.

\bibitem{Il}
Iliev B.Z., Handbook of normal frames and coordinates, \href{http://dx.doi.org/10.1007/978-3-7643-7619-2}{\textit{Progress in
  Mathematical Physics}}, Vol.~42, Birkh\"auser Verlag, Basel, 2006.

\bibitem{KT}
Kijowski J., Tulczyjew W.M., A symplectic framework for f\/ield theories,
  \href{http://dx.doi.org/978-3-540-35016-3}{\textit{Lecture Notes in Phys.}}, Vol.~107, Springer-Verlag, Berlin~-- New
  York, 1979.

\bibitem{La}
Landsman N.P., Mathematical topics between classical and quantum mechanics,
  \href{http://dx.doi.org/10.1007/978-1-4612-1680-3}{\textit{Springer Monographs in Mathematics}}, Springer-Verlag, New York, 1998.

\bibitem{LP}
Lauda A.D., Pfeif\/fer H., Open-closed strings: two-dimensional extended {TQFT}s
  and {F}robenius algebras, \href{http://dx.doi.org/10.1016/j.topol.2007.11.005}{\textit{Topology Appl.}} \textbf{155} (2008),
  623--666, \href{http://arxiv.org/abs/math.AT/0510664}{math.AT/0510664}.

\bibitem{Mi}
Milnor J.W., Topology from the dif\/ferentiable viewpoint, Based on notes by
  David W.~Weaver, The University Press of Virginia, Charlottesville, Va.,
  1965.

\bibitem{Mo}
Moser J., On the volume elements on a manifold, \href{http://dx.doi.org/10.1090/S0002-9947-1965-0182927-5}{\textit{Trans. Amer. Math.
  Soc.}} \textbf{120} (1965), 286--294.

\bibitem{O3}
Oeckl R., General boundary quantum f\/ield theory: foundations and probability
  interpretation, \href{http://dx.doi.org/10.4310/ATMP.2008.v12.n2.a3}{\textit{Adv. Theor. Math. Phys.}} \textbf{12} (2008),
  319--352, \href{http://arxiv.org/abs/hep-th/0509122}{hep-th/0509122}.

\bibitem{O2}
Oeckl R., Two-dimensional quantum {Y}ang--{M}ills theory with corners,
  \href{http://dx.doi.org/10.1088/1751-8113/41/13/135401}{\textit{J.~Phys.~A: Math. Theor.}} \textbf{41} (2008), 135401, 20~pages,
  \href{http://arxiv.org/abs/hep-th/0608218}{hep-th/0608218}.

\bibitem{O1}
Oeckl R., Af\/f\/ine holomorphic quantization, \href{http://dx.doi.org/10.1016/j.geomphys.2012.02.001}{\textit{J.~Geom. Phys.}} \textbf{62}
  (2012), 1373--1396, \href{http://arxiv.org/abs/1104.5527}{arXiv:1104.5527}.

\bibitem{O}
Oeckl R., Holomorphic quantization of linear f\/ield theory in the general
  boundary formulation, \href{http://dx.doi.org/10.3842/SIGMA.2012.050}{\textit{SIGMA}} \textbf{8} (2012), 050, 31~pages,
  \href{http://arxiv.org/abs/1009.5615}{arXiv:1009.5615}.

\bibitem{Sa}
Sati H., Duality and cohomology in {$M$}-theory with boundary, \href{http://dx.doi.org/10.1016/j.geomphys.2011.11.012}{\textit{J.~Geom.
  Phys.}} \textbf{62} (2012), 1284--1297, \href{http://arxiv.org/abs/1012.4495}{arXiv:1012.4495}.

\bibitem{Sc}
Schwarz G., Hodge decomposition~-- a method for solving boundary value
  problems, \href{http://dx.doi.org/10.1007/BFb0095978}{\textit{Lecture Notes in Math.}}, Vol.~1607, Springer-Verlag,
  Berlin, 1995.

\bibitem{Seg:cftdef}
Segal G., The def\/inition of conformal f\/ield theory, in Topology, Geometry and Quantum Field Theory,
Cambridge University Press, Cambridge, 2004, 421--577.

\bibitem{We}
Weinstein A., Symplectic categories, \href{http://dx.doi.org/10.4171/PM/1866}{\textit{Port. Math.}} \textbf{67} (2010),
  261--278, \href{http://arxiv.org/abs/0911.4133}{arXiv:0911.4133}.

\bibitem{Wi}
Witten E., Two-dimensional gauge theories revisited, \href{http://dx.doi.org/10.1016/0393-0440(92)90034-X}{\textit{J.~Geom. Phys.}}
  \textbf{9} (1992), 303--368, \href{http://arxiv.org/abs/hep-th/9204083}{hep-th/9204083}.

\bibitem{Wo}
Woodhouse N.M.J., Geometric quantization, 2nd ed., \textit{Oxford Mathematical
  Monographs}, The Clarendon Press, Oxford University Press, New York, 1992.

\end{thebibliography}
\end{document}